%% file: 00manuscript.tex
\documentclass[manuscript]{acmart}

\AtBeginDocument{%
  \providecommand\BibTeX{{%
    \normalfont B\kern-0.5em{\scshape i\kern-0.25em b}\kern-0.8em\TeX}}}

\setcopyright{acmcopyright}
\copyrightyear{2020}
\acmYear{2020}
\acmDOI{}

\input{macro}
\usepackage{breakurl}

\begin{document}



\title{``A cold, technical decision-maker'': Can AI provide explainability, negotiability, and humanity?}




\author{Allison Woodruff}\affiliation{\institution{Google}}\email{woodruff@google.com}
\author{Yasmin Asare Anderson}\affiliation{\institution{Ipsos}}\email{yas.asareanderson@ipsos.com}
\author{Katherine Jameson Armstrong}\affiliation{\institution{Ipsos}}\email{katherine.j.armstrong@ipsos.com}
\author{Marina Gkiza}\affiliation{\institution{Ipsos}}\email{marina.gkiza@ipsos.com}
\author{Jay Jennings}\affiliation{\institution{PixelInsight}}\email{jay@pixelinsight.co.uk}
\author{Christopher Moessner}\affiliation{\institution{Ipsos}}\email{christopher.moessner@ipsos.com}
\author{Fernanda Viégas}\affiliation{\institution{Google}}\email{viegas@google.com}
\author{Martin Wattenberg}\affiliation{\institution{Google}}\email{wattenberg@google.com}
\author{Lynette Webb}\affiliation{\institution{Google}}\email{lwebb@google.com}
\author{Fabian Wrede}\affiliation{\institution{Google}}\email{fwrede@google.com}
\author{Patrick Gage Kelley}\affiliation{\institution{Google}}\email{patrickgage@acm.org}

\renewcommand{\shortauthors}{Woodruff, et al.}

    
\begin{abstract}
Algorithmic systems are increasingly deployed to make decisions in many areas of people’s lives. The shift from human to algorithmic decision-making has been accompanied by concern about potentially opaque decisions that are not aligned with social values, as well as proposed remedies such as explainability. We present results of a qualitative study of algorithmic decision-making, comprised of five workshops conducted with a total of 60 participants in Finland, Germany, the United Kingdom, and the United States. We invited participants to reason about decision-making qualities such as explainability and accuracy in a variety of domains. Participants viewed AI as a decision-maker that follows rigid criteria and performs mechanical tasks well, but is largely incapable of subjective or morally complex judgments. We discuss participants' consideration of humanity in decision-making, and introduce the concept of `negotiability', the ability to go beyond formal criteria and work flexibly around the system.
\end{abstract}




\maketitle


\section{Introduction}
\input{01intro}

\section{Background}
\input{03background}

\section{Methodology}
\input{05method}

\section{Findings}
\label{sec:findings}
\input{07findings}

\section{Discussion}
\input{09discussion}

\section{Conclusion}
\input{11conclusion}



\sloppy
\bibliographystyle{ACM-Reference-Format}
\bibliography{AIAP}


\clearpage
\begin{appendices}
\section{Scenario Methodology}

\input{figtab/methodtab}

\end{appendices}

\typeout{get arXiv to do 4 passes: Label(s) may have changed. Rerun}
\end{document}

%% file: macro.tex
\usepackage[title]{appendix}

\usepackage{colortbl}
\definecolor{ltgray}{RGB}{175,175,175}
\definecolor{darkplum}{RGB}{80,30,60}

\newcommand\aff[1]{\textcolor{darkplum}{\emph{--#1}}}
\newcommand\affpre[1]{\textcolor{darkplum}{\emph{#1}}}

\newenvironment{listquote}
{ \setlength{\partopsep}{4pt}
  
  \begin{itemize}
    \fontsize{8.2pt}{8.2pt}\selectfont
    \setlength{\itemsep}{1pt}
    \setlength{\parskip}{6pt}
    \setlength{\parsep}{0pt}  
                                 }
{  
  \end{itemize}                  } 

\newcommand\fixme[1]{}

%% file: 01intro.tex
Motivated by expectations of improved efficiency, efficacy, consistency, and objectivity, algorithmic systems are increasingly deployed to make decisions in many areas of people’s lives, from finance, employment, medicine and criminal justice to advertising, media recommendations and more~\cite{citron2007,edwards2017slave,pasquale2015,selbst2018intuitive}. This shift from individual or organizational human decision-making to algorithmic decision-making has been accompanied by concern that important decisions are being made in new ways that are often mysterious, and that current oversight does not adequately ensure social values such as fairness, privacy, and safety~\cite{citron2014,kroll2016,pasquale2015,selbst2018intuitive}.

A variety of approaches have been proposed in technical, social science, legal, and regulatory communities to address this issue and provide accountability for algorithmic systems. However, the best path forward is as yet unclear. For example, increased transparency or explainability are often proposed as mechanisms to provide accountability, and frequently feature in regulatory and tech ethics requirements or discussions~\cite{fjeld2020,goodman2017european,highlevel2019,kaminski2019right,wachter2017}. While these have significant potential upside, concerns have been raised that such mechanisms are unlikely to be sufficiently effective or satisfactory, due to fundamental challenges in producing meaningful human explanations of the operation of complex machines~\cite{ananny2016,edwards2017slave,kroll2016,sendak2020}. Further, in some contexts practical constraints may require that transparency or explainability be traded off against qualities such as a system’s accuracy or its ability to resist adversarial attacks~\cite{burrell2016machine,kroll2016,london2019}.

Limited understanding of public attitudes towards algorithmic decision-making makes it difficult to include public needs and desires in balancing these complex tradeoffs. To inform policy and technical approaches to algorithmic decision-making, we present results from a qualitative study comprised of five workshops conducted with a total of 60 participants in Finland, Germany, the United Kingdom, and the United States. Our contributions are as follows:

\begin{itemize}
\item We present results from a qualitative study of algorithmic decision-making conducted in four countries
\item We identify key desirable qualities of decision-makers and discuss how AI is perceived as able or not able to provide them
\item We discuss how algorithmic decision-making is intended to regularize institutional processes and make them more efficient, but in so doing is perceived as reducing flexibility, compassion, and other qualities desired by recipients of a decision
\item We consider how our findings inform potential design of algorithmic decision-making systems, as well as allocation of responsibilities between AI and humans in human-in-the-loop approaches
\end{itemize}

In the remainder of the paper, we review relevant background, describe our methodology, present and discuss our findings, and conclude.

%% file: 03background.tex
\subsection{Algorithmic Decision-Making Systems, Artificial Intelligence, and Machine Learning}
In recent decades, physical and digital systems across a range of sectors have increasingly transitioned to automation for a range of functions including decision-making,\footnote{In this article we use the term algorithmic decision-making systems, following~\cite{diakopoulos2016,kaminski2019binary} and others. The term automated decision-making is often used synonymously, commonly describing decision-making functions that are carried out in part or in full by a device or machine~\cite{parasuraman2000}.}
in hopes of improving efficiency and performance, and regularizing decisions by removing human and procedural variability~\cite{eubanks2017,immonen2018,ostrom2002}.
The significant impact, risks, and challenges of this transition have been an object of substantial study, raising concerns related to fairness, privacy, safety, and more as algorithmic systems increasingly shape information people are exposed to as well as influence decisions about employment, finances, and other opportunities~\cite{barocas2016,boyd2012,citron2014,cottom2015,eubanks2017,executive2016,fourcade2013,friedman1996,ftc2016,gillespie2014relevance,introna2000,lustig2016,oneil2016,pasquale2015,sandvig2015}.

Artificial intelligence (AI) underlies many algorithmic decision-making systems. 
Machine learning is a particular type of artificial intelligence that has the ability to learn and improve over time, and it is an increasingly common choice for many applications including decision-making.
Accordingly, in our study we focused primarily although not exclusively on algorithmic decision-making systems that use different types of AI, including machine learning.

\subsection{Accountability, Transparency, Explainability, and Interpretability}
In search of mechanisms to detect and address social ills that may result from the use of algorithmic decision-making systems, and building on the expectation that insight into a system's production, operation, and effects can enable accountability, many regulators, civil society groups, and scholars have called for a cluster of related ideas encompassing transparency, explainability, and interpretability.\footnote{These notions are strongly related and these terms are often used interchangeably although some scholars and disciplines also distinguish their meanings in various ways ~\cite{biran2017,gilpin2018,kluttz2019,lipton2016,miller2019,rader2018}.
For our purposes in this article, we define transparency as broad visibility into the construction, behavior, and processes of a system; explainability as the ability to provide human understandable reasons for individual decisions or a set of decisions made by an algorithmic system; and interpretability as the ability of a given model to explain how it calculates specific decisions.}
Consideration of these approaches dates to the 1970s with the recognition of the need to explain the decisions of expert systems, with a resurgence of this topic in recent years as contemporary machine learning models have become both pervasive and increasingly complex~\cite{abdul2018,biran2017,gilpin2018,miller2019,selbst2018intuitive}.
Due to its often impenetrable and opaque nature, machine learning, particularly deep learning, has generated concern and attracted heavy critique as it can be difficult to interrogate its reasoning and remediate potential injustices or errors ~\cite{burrell2016machine,gillespie2014relevance}.
Interpretability is therefore a topic of substantial debate at the highest levels of machine learning expertise~\cite{caruana2017,rudin2019},
with many arguing there is an inherent tension between a system's accuracy and its interpretability, but others arguing it is possible
to finesse this technical challenge with methods such as creating human-understandable models that are proxies for more complex models, intentionally designing models that are more readily interpretable, or conducting ``outside the box'' analyses of system design and behavior~\cite{breiman2001,gilpin2018,gupta2016,lipton2016,ribeiro2016,selbst2018intuitive}.
A number of scholars further argue that
making systems more understandable is at odds with other values such as
protecting systems from adversarial attacks or seamless design~\cite{burrell2016machine, eslami2019, kroll2016},
and even when there is not an inherent tradeoff, it can distract from more important priorities such as empirical (non-explanatory) verification of system performance~\cite{ananny2016,edwards2017slave,lipton2016,london2019,sendak2020}.
More fundamentally, there is a growing recognition that transparency, explainability, and interpretability are often intended as a means to an end, for example laying the groundwork to challenge a decision or to ensure fairness, yet they provide inadequate support for such underlying objectives~\cite{kluttz2019,rader2018,woodruff2019}.
As a result, a number of approaches have been articulated in recent years that hew more closely to direct needs of users and society. 
Counterfactuals specify conditions under which a different decision would have been made~\cite{karimi2020,russell2020,wachter2018counterfactual}. Contestability provides the ability to challenge decisions, and allows domain experts to collaborate with and correct predictive algorithms~\cite{hirsch2017,kluttz2019,vaccaro2019}. Recourse considers the ability to correct faulty or incomplete data, and also identify clear actions an individual can take in the future to get a different outcome~\cite{ustun2019,venkatasubramanian2020}.

\subsection{Algorithmic Literacy and Explanation Design}

Beyond the challenges described above, provision of good explanations is further complicated by the fact that end users often have fundamental questions or misconceptions about the operation of algorithmic systems~\cite{bucher2017algorithmic, eslami2015always, rader2015understanding, ur2012smart, warshaw2016intuitions}, and the construction of useful explanations remains elusive~\cite{nunes2017,rader2018}. These challenges notwithstanding, a number of studies have shown that explanations have potential to increase users’ trust and satisfaction~\cite{biran2017,nunes2017}, and they remain an object of great interest with a number of efforts pursuing best practices for their design~\cite{balog2020, cai2019, eslami2016, herman2017, miller2019, nunes2017, poursabzi2018, wang2019}.

\subsection{Public Perception of AI and Algorithmic Decision-Making}

A number of studies have explored public perception of AI, for example, survey research~\cite{arm2017, blumberg2019, cave2019, edelman2019, ipsos2019, kelley2020, mozilla2019, northeastern2018, west2018brookings, zhang2019artificial}, sentiment analysis~\cite{fast2017long, garvey2019sentiment}, and narrative analysis~\cite{cave2018portrayals}, most often in Western, English-speaking contexts.
AI is viewed as likely to have significant impact on the future and overall expectation is often positive. At the same time, AI is neither interpreted as exclusively beneficial nor exclusively disadvantageous, and public response is often ambivalent~\cite{arm2017,blumberg2019,kelley2020}.
Specific concerns have been expressed regarding social issues, such as privacy, job loss, AI benefiting the wealthy and harming the poor, and AI increasing social isolation and reducing human capabilities~\cite{edelman2019,kelley2020}.
Resonating with this interest in the impact of AI on social issues, 
some public perception research has focused specifically on its use in decision-making, e.g.
a survey in the United Kingdom found that people were not very familiar with the use of automated decision systems, and indicated low levels of support and high levels of opposition to them~\cite{balaram2018}.

\subsection{Workshop as Method}
In taking up a workshop format, we draw on traditions within and just beyond HCI. This includes programs of participatory action research, participatory design, and living labs~\cite{disalvo2008, disalvo2011, jungk1987, rosner2016}. Within the context of HCI and design research, workshop approaches often seek to invite members of the public to engage with practices of design while exploring values and beliefs around technology with each other, positing alternative techniques and outcomes. We also draw inspiration from Citizens' Juries, a form of deliberative democracy dating from the 1970s that brings together individuals from different backgrounds to spend several days hearing expert evidence and considering a public policy question~\cite{armour1995, balaram2018, ico2019, jefferson2004, oswald2019}. 
Two deliberative democracy studies run in the United Kingdom that used Citizens' Juries methodology are closest in nature to our own work.
One explored under what circumstances participants would find the use of automated decision-making appropriate, finding they valued accountability and ethical development processes~\cite{singh2019}.
Another focused on how participants prioritized accuracy versus explainability for AI decision-making, reporting that context mattered and accuracy outweighed explainability in healthcare but not other non-medical situations presented~\cite{oswald2019,singh2019}.
We build on, extend, and broaden this previous work by exploring complementary research objectives; exploring a wider range of domains, tradeoffs, and types of transparency; and expanding the scope to three additional countries beyond the United Kingdom.

%% file: 05method.tex
In order to better understand perception of algorithmic decision-making systems, 
we partnered with our co-authors from [\emph{anonymized market research firm}]
to conduct five workshops with 60 participants in major cities in Finland, Germany, the United Kingdom, and the United States. These countries were selected due to European and United States regulatory interest in accountability of AI systems~\cite{highlevel2019, kaminski2019right} and pre-existing literature on perceptions of AI or algorithmic decision-making that we could build on and triangulate with, for example~\cite{balaram2018, cave2019, ico2019, kelley2020, zhang2019artificial}.
Partnering with an international market research firm ensured professional quality recruiting and moderation across multiple countries, languages, and cultural contexts. Our research objectives were to learn more about the following:

\begin{itemize}
\item{What qualities do members of the public want in algorithmic decision-making systems?}
\item{To what extent do members of the public value or prioritize explainability in particular, for what reasons?}
\item{What qualities do members of the public believe an AI-based algorithmic decision-making system has, or is capable of? In what circumstances do they see its use as beneficial?}
\end{itemize}

\subsection{Participants}
We recruited 12 participants for each of five workshops (two in Germany and one each in the other countries) through professional recruiting firms in each country. In each region, participants were recruited to represent a diverse spread across the following: age, gender, income, life stage, socioeconomic status, educational level, digital literacy, and exposure to AI. They were recruited to represent a mix of ethnicity representative of the region. Participants were compensated for their time, at or above the living wage for their area.
\subsection{Workshops}
Each group participated in a 3-hour workshop at third party facilities in each country during November and December 2019. Two moderators were in the discussion rooms with the 12 participants, and participants were aware that several researchers and staff were in observation rooms or viewing remotely. All moderators were from a professional qualitative research background and were fluent in the local language. In non-English speaking countries (Finland and Germany), simultaneous translators were present in the observation rooms and their translations were broadcast to remote viewers and captured on recordings so that the researchers and staff observing the sessions could hear the discussion and pass questions and input to the moderators in real-time. Both moderators and simultaneous translators received specific training in the objectives and content of the workshops before the sessions. Refreshments were available throughout the sessions. Workshops followed this agenda (times are approximate):
\begin{itemize}
\item{Introductions and warm-up activity sharing experiences with human decision-making -- \emph{10 minutes}}
\item{Presentation about AI and machine learning, their use in decision-making, tradeoffs, and Q/A -- \emph{60 minutes}}
\item{First scenario (in two breakout groups, with six participants each) -- \emph{40 minutes}}
\item{Break -- \emph{10 minutes}}
\item{Second scenario (in same breakout groups as the first scenario) -- \emph{30 minutes}}
\item{Reconvene and summarize -- \emph{30 minutes}}
\end{itemize}


The scenarios described hypothetical situations in which organizations were considering implementing algorithmic decision-making and were choosing among three systems with different characteristics (see Tables~\ref{tab:scenarios} and~\ref{tab:systems} in Appendix A). For each scenario, we began by sharing a handout describing it and asking participants to take a few minutes to read and consider it. We then invited participants to share their initial reactions and facilitated a group discussion of the scenario. Midway through the discussion, we asked participants to fill out a handout and then vote in group discussion on which system they preferred.\footnote{While most of the procedure was consistent across the different countries, the voting process was somewhat variable as the study evolved, for example, in some cases participants voted both midway and at the end of the discussion.
As a result, we do not report voting data here, although it was consistent with the qualitative findings we report.}
The scenarios we discussed represented a wide range of domains and we encouraged discussion of other domains, so the discussion often branched out to other areas in which algorithmic decision-making might be used. After completing both scenarios, all participants joined back together and we concluded the workshop with a broad group discussion where participants reflected on ideas that had emerged throughout the session and synthesized and summarized their opinions. If participants requested, they were informed of the sponsor of the research at the end of the session. During the workshops, we took care to encourage collaborative interpretation, problem-solving, and discussion among participants, and to make space for all participants to share their ideas and opinions.

After each session, moderators, researchers, and other team members held a debriefing session to discuss early observations and themes, as well as potential procedural refinements for future sessions such as timing adjustments. Detailed notes were taken on observations and themes for each country as the study progressed. All sessions were video- and audio-recorded and transcribed in English. For the United Kingdom and the United States, the transcripts were then compared against the original recordings for accuracy, and for Finnish and German the transcripts were reviewed and compared against the simultaneous translations for accuracy.

\subsection{Material Development and Translation}
We developed materials designed to explore our research objectives, drawing heavily for inspiration on Citizens’ Juries work previously conducted in the United Kingdom~\cite{balaram2018, ico2019, oswald2019, singh2019}.
This included a slide deck and accompanying script to provide participants relevant background on AI and machine learning; descriptions of rule-based systems, conventional machine learning, and deep learning; the use of AI and machine learning in decision-making; and tradeoffs between accuracy and various forms of transparency. In some sessions we also showed a short introductory video about machine learning available on the internet.
We also developed four hypothetical scenarios chosen to represent decision-making in topical and diverse contexts: stroke diagnosis, job applicant screening, credit scoring, and law enforcement, as shown in Table~\ref{tab:scenarios} (Appendix A).
For each scenario, participants were presented with three different systems which varied in a number of ways, similar to~\cite{ico2019,oswald2019}, as shown in Table~\ref{tab:systems} (Appendix A). The primary differences centered on accuracy versus a form of transparency (explainability, appeal, or oversight), which we emphasized due to the centrality of this discussion in current policy and academic discussions. In some cases the magnitude and nature of the tradeoff were exaggerated in order to prompt useful discussion, rather than strictly adhering to specific current performance levels, again similar to~\cite{ico2019,oswald2019}.
Finally, we developed a moderator discussion guide to use with these materials. The materials were developed in English, and after they were complete, our market research partner translated materials into Finnish and German.


\subsection{Analysis}
In our analysis, we used a general inductive approach~\cite{thomas2006inductive}, which relies on detailed readings of raw data to derive themes relevant to evaluation objectives. In our case, the primary evaluation objectives were those listed at the beginning of this section, chosen to inform technical and policy approaches to algorithmic decision-making. After the workshops were over, one member of our research team reviewed all transcripts and prepared participant precis and breakout group precis summarizing the main points for each individual and breakout group, and then prepared a spreadsheet summarizing the key ideas raised by each participant and breakout group. This was work done in active consultation with another member of the research team and built on the collective debriefs conducted by the broader team after each workshop. Throughout, we continued to identify emergent themes~\cite{beyer1997contextual}, iteratively revising and refining categories. In keeping with the general inductive approach, our analytic process yielded a small number of summary categories. We summarize those most relevant to our evaluation objectives in Section~\ref{sec:findings}.


\subsection{Limitations}
We note several limitations of our methodology that should be considered when interpreting this work. First, it carries with it the standard issues attendant with qualitative methodologies and group interviews. While it yields rich insight, one must be cautious in generalizing beyond those studied or drawing conclusions across countries~\cite{armour1995}. 
Second, we asked participants to consider hypothetical scenarios that most did not perceive as common in their daily lives at the current time. They may have had different reactions if they had been speaking to commonplace experiences, rather than extrapolating from more limited experiences in their current lives.
Third, we studied a small number of scenarios, often with high stakes import. Different scenarios might yield varying results or further refine the insights we share here.
Fourth, while we were heavily inspired by the Citizens’ Juries approach~\cite{armour1995, jefferson2004}, our workshops did not follow all aspects of this approach, most notably in that they were significantly shorter and more lightweight engagements (three hours in our case, versus several days for Citizens’ Juries) with less instructional content, no expert witnesses, and less discussion. In this way, while our data represents participants who were provided substantial information, our data does not represent what members of the public might think with more extensive intervention. 
Fifth, the discussion allowed open-ended conversation but also at some point required participants to express a clear choice for each scenario based largely on the tradeoff between explainability and accuracy. While Citizens’ Juries typically require jurors to make such a clear choice which prompts useful insight and discussion~\cite{oswald2019}, it also focuses some of the discussion on particular issues. To mitigate this, both in moderation and analysis, we took a wide view beyond these specific choices.
Sixth, while members of the research team and/or market research partner team have experience conducting research in all markets studied, most reside in the United States or the United Kingdom. We have worked to minimize the risk of misinterpretation by collaboration and discussion with in-country colleagues but recognize that our interpretations may lack context or nuance that would have been more readily available to local residents.

%% file: 07findings.tex
In this section, we describe the main findings that emerged from our analysis. We begin by describing how participants perceived algorithmic decision-making systems as offering efficiency and performance advantages in some circumstances but potentially making substantial errors in others, as well posing a potential threat to dignity and autonomy. We discuss the perceived opportunity for explanations and human intervention to ameliorate some of these concerns. We then summarize the desirable qualities of decision-makers, whether AI or human, and close this section by discussing situational factors that influence tradeoffs among those qualities and influence the allocation of responsibility to algorithmic decision-making systems versus humans.

\subsection{AI as an emotionless, mechanical decision-maker}
\label{sec:technical}

Participants recognized strong advantages of algorithmic decision-making in some circumstances, such as speed, scale, or improved accuracy. 
Participants perceived AI as performing objective, mechanical tasks well, and characterized it as thinking in black-and-white terms and following rigid criteria. Correspondingly, they often viewed AI as incapable of subjective or morally complex judgments.\footnote{Our discussions with participants focused on AI-based algorithmic decision-making systems, with a broad range of functionality from basic rule-based systems to advanced deep learning systems. Accordingly, when participants speak of AI they typically refer to it in the context of an AI-based algorithmic decision-making system.}

\begin{listquote}
\item ``We are talking about 3 systems that have no gut instinct algorithm or anything similar, but which simply make decisions based on clear structures.'' ~\aff{Germany A08}~\footnote{For ease of reading, we have followed editing conventions consistent with applied social science research practices as described in~\cite{corden2006verbatim}. Specifically, we edited quotes to remove content such as filler words and false starts, and in some cases we re-punctuated. We use ellipses to indicate substantial omissions. We identify each participant by country name plus a numerical identifier from 01 to 12 for each workshop. For Germany, we distinguish between the two workshops by prepending A or B to the numerical identifier.}
\item \affpre{Germany A11}: ``This is all theory, and life is everything in between! Artificial Intelligence has black or white. I mean, when it comes to the rules. And that's where we also, as people, can make different decisions based on the situation.'' \newline
\affpre{Germany A07}: ``We've heard about a system which, in the personnel field, can do the work for us and decide within two or three minutes whether the applicant is good or not... But then there's a certain `gut feeling' involved, there's not just black or white or grey, but also something in between, and you have to look at whether somebody as a whole actually fits, and the system can't do that, because it only looks at `studies' yes or no, `foreign languages' yes or no, but there are always some things in between that a computer, from my perspective, can't evaluate.''
\end{listquote}

Beyond perceiving AI as well-suited to making straightforward decisions, participants sometimes saw AI as appropriate for complex technical tasks, such as autopilot for airplanes or cars, with some variation in preferred tradeoffs between safety (accuracy) versus explanation.

\begin{listquote}
\item ``I would imagine it could apply to areas of autonomous driving or flying or similar. Complex things that are technically-based where I, as a user, would like to have the highest amount of safety [and would not require explanation].''~\aff{Germany A09}
\item ``I wanted to add something regarding trust. For me, it depends on how difficult the task is... When I know the task is quite simple and it has received much data, then I think, I don't care how it does it, but okay, it has done it. But when it is a more complex task, then I would rather like to know how it works. Otherwise I wouldn't feel too safe.''~\aff{Germany B05}
\end{listquote}

By contrast, AI was seen as poorly suited to tasks that dealt with non-specific factors, such as complex moral judgments or assessments of intangible factors, consistent with an argument made in legal scholarship that policies that require the exercise of human discretion cannot be automated~\cite{citron2007}. Participants in a study in the United Kingdom believed that machines operate with logic and rules that cannot form the basis of ethical decisions~\cite{ipsosmori2017}; here we develop this argument further and provide evidence of it in other countries.

\begin{listquote}
\item ``{[AI]} doesn't allow for those things that can't be defined by specifics.''~\aff{United Kingdom 12}
\item ``In these kinds of humane decision-making things, where more complicated, abstract things are weighed… justice, fairness and such… something much higher than just crunching numbers, for that artificial intelligence is not suitable.''~\aff{Finland 11}
\item ``I'd say it depends on the role ethics play in the situation. So, when it comes to hospital decisions, we very likely all agree that we want to save lives, everybody should be as healthy as possible and therefore, if technology can help us with that, we might be happy to take it. Whereas with other questions, where we are not even sure how we should act as humans, questions where we don't even agree on how much power the police should actually have, in those cases AI might be a tool that some reject as a matter of principle. And these are no longer yes or no questions, they are far more complicated because they are ethical, legal and philosophical. And when even we as humans can't even agree on that, there's no point in bringing AI into it.''~\aff{Germany B04}
\end{listquote}

Many participants emphasized that AI does not take into account soft factors (qualities that are intangible or difficult to specify), leading to strong skepticism that AI could make good or fair decisions for tasks where such factors are important, consistent with~\cite{singh2019}.

\begin{listquote}
\item ``If I was an applicant myself, then in my opinion, it would be unfair for me if only my knowledge was evaluated, and not that `that's a really good person, matches this team exactly, it's exactly who I'm looking for.' Can it do that, the artificial intelligence, can it know what type of people I like and with whom I like to work? ... I don't believe an artificial intelligence can know whether I'm suitable for some job or not.''~\aff{Finland 04}
\item \affpre{Germany B11}: ``{[Even with a hip replacement]}, there are still certain things to weigh... there are still soft factors involved and a human assessment of whether this is still justified for someone.'' \newline
... \newline
\affpre{Germany B01}: ``Even if the computer says you need a new hip, it always depends on how much a person is suffering. The individual physical strain, how the body works, something does not fit together, whether you got an inflammation. This is individually different. It always depends on the individual level of suffering and the future intelligence cannot decide this at all.'' \newline
... \newline
\affpre{Germany B07}: ``Many also managed to overcome cancer based on their personal willpower. A machine won't be able to tell how strong a human's will is.''
\end{listquote}

AI was sometimes characterized as cold and emotionless, consistent with~\cite{balaram2018,kelley2020}, and lacking human qualities such as `gut instinct.'

\begin{listquote}
\item \affpre{Finland 04}: ``I'm thinking about whether it becomes so-called too intelligent, that it becomes a cold, technical decision-maker... Then certain issues, where there should be the humane aspect, can it then do such things?'' \newline
... \newline
\affpre{Finland 06}: ``I was reminded of the old saying: `Good servant but poor master.'{}''
\end{listquote}

\subsection{Mysterious Dictates}
\label{sec:mysterious}
Concerns have long been raised in both scholarly and fictional literature that algorithmic decision-making and automated systems can undermine human autonomy if powerful institutions issue consequential decisions that cannot be understood or appealed~\cite{cave2019,edwards2017slave,kaminski2019binary,west2018censored,selbst2018intuitive}. Autonomy is considered essential to human wellbeing, making the notion of such unfettered power both frightening and dehumanizing~\cite{cave2018portrayals,eubanks2017,kelley2020,kluttz2019,selbst2018intuitive,venkatasubramanian2020}. Participants often brought up these issues. For example, the notion that deep learning systems learn on their own, developing their own unseen rules and assumptions, was sometimes experienced as disturbing, eerie or even scary.

\begin{listquote}
\item ``Deep learning concerns me because it seems totally computer led and based, and there's no human intervention. Although they all have ultimately some human behind it... But I think the deep-learning route is just making up its own set of rules and assumptions.''~\aff{United Kingdom 12}
\end{listquote}

 Participants associated this autonomy and opacity with a loss of human influence and control. Sometimes this was framed as a transfer of power from humans to AI, or even as humans being ``at the mercy of the machines.'' In some cases these concerns resonated with common media tropes about AI taking over society.

\begin{listquote}
\item ``It is difficult, if there is something starting to develop itself, and you can't influence it anymore, at least only in a limited way.''~\aff{Germany B09}
\item ``I also got a scary feeling from {[deep learning]}, the feeling that we've given power to somewhere else.''~\aff{Finland~07}
\item \affpre{Germany B02}: ``[System C is] not transparent. It can't give a reason.'' \newline
\affpre{Germany B11}: ``As they don't know it themselves. Nobody knows why the computer is making a certain decision because it is learning by itself, somewhere deep inside.'' \newline
\affpre{Germany B07}: ``And we are supposed to trust them. [laughter]'' \newline
\affpre{Germany B02}: ``... there is no argument against the decision.'' \newline
\affpre{Germany B10}: ``So you elevate the machine to be equal to God.''
\end{listquote}

Some participants formulated the use of deep learning in algorithmic decision-making as overreliance on machines, expressing concern about blindly accepting decisions made by machines. Some were also concerned about humans subjecting themselves to the dictates of algorithmic decision-making systems, saying for example that a deep learning credit scoring system that did not provide explanations was ``too authoritarian.'' 

\begin{listquote}
\item ``You don't want somebody to just dictate everything for you. You want to have the ability to be like... `Wait, that's not really right.'{}''~\aff{United States 07}
\item ``I'm so anti-relying on machines and this would force me to say, `... because you, the machine tell me this is 95\% accurate, I have to automatically accept it.' ... `Well, F you, sorry, I don't agree with it.'{}''~\aff{United States 12}
\end{listquote}

Such concerns were exacerbated by fear or anger at the thought of ending up in the group about whom AI made an inaccurate decision, and a sense that it was ``random'' chance whether one ended up in that group. Participants sometimes compared subjecting oneself to decisions from such a system to gambling at poker or dice. A few participants also worried that it would be difficult to escape from an inaccurate decision once it had been made, as decisions might be shared across institutions, leaving them essentially powerless and without recourse.

\begin{listquote}
\item ``If I'm one of that 5\%, then it's really shitty that I'm accused of something I really haven't done.''~\aff{Finland 01}
\item ``Transparency {[is more important than accuracy]}. Because 95\% means that there are still five percent which are wrong. And if I am part of the 5\% then it is bad.''~\aff{Germany A03}
\item ``It only sucks if you are the five percent. That's the issue... for most it is good, but for those for whom it does not work, it is shit... Oh man, that's like playing poker. [laughter] That is so much like playing poker. Even with the best hand, you can always lose.''~\aff{Germany B11}
\item\affpre{Germany A09}: ``I think this would be bullshit [to not get an explanation regarding a decision, for example, the reason your application for a loan has been rejected], because you cannot escape from this hamster's wheel. No matter what you do, you cannot escape it. You are simply unlucky. And you cannot learn from it. You cannot get behind the reason et cetera, you cannot do anything.'' \newline
~\affpre{Germany A05}: ``You cannot prepare better for the next job application because you don't know why they did not accept your previous application.'' \newline
~\affpre{Germany A07}: ``And if everybody works with the same AI system, it would not matter whether I write a hundred or a thousand applications. The AI system would always say, `You are out.' ... Nowadays, the situation is different. I can go to [three different institutions] and they might apply different criteria... With System C I have to hope that I don't belong to the 5\% group.'' \newline
~\affpre{Germany A09}: ``Yes... There will always be winners or losers, however, once you are a loser, you will always be a loser.''
\end{listquote}

\subsection{What do explanations offer?}
\label{sec:explanations}

Explanations are a potential mechanism to ameliorate the concerns raised in the previous sections, and participants portrayed explanations as having a number of useful or desirable functions as summarized in Table~\ref{tab:purposes}. Explanations were seen as providing transparency, allowing people to detect errors (e.g. incorrect data leading to a poor credit score), and laying the groundwork for appeals to change decisions. Explanations were also seen as a mechanism for oversight and accountability, for example, a way to detect and expose discrimination in job application screening decisions.

\input{figtab/purposes}

\begin{listquote}
\item \affpre{United States 02}: ``You need the explanation to make the appeal.'' \newline
\affpre{United States 01}: ``First. That's the foundation which you base your appeal on.'' \newline
... \newline
\affpre{United States 05}: ``The explanation... that's the most important component of it. If you get the explanation, you might realize that there is no reason for you to appeal or you might not need to appeal.''
\end{listquote}

Beyond corrective measures and transparency, explanations were highlighted as informing personal future behavior or actions, for example, understanding how to modify one's future financial decisions in order to increase a credit score.

\begin{listquote}
\item \affpre{United States 12}: ``What's bothering me about deep learning is that there's no way of explaining how it reaches its decision. That is bothering me... because, when push comes to shove, I want to know, `How'd you reach the decision? What was the basis? What was the factors?' I need to know.'' \newline
\affpre{United States 06}: ``It's going to be a common problem that somebody has a low credit score, and if there's no way to tell them how to improve, that's a little frustrating.''
\item ``Sometimes, there's not a need for an appeal, but there's a need for improvement. So if you have help on what you could do to improve, then that's the most important.''~\aff{United States 05}
\item ``If you had a system where they could pick out specific bits of the CV... what it would allow people to do is to go away and go, `Oh right, okay. I just haven't got this job because I haven't got the criteria. This is the specific things that I need to work on.'{}''~\aff{United Kingdom~08}
\item ``What can I do myself to better my situation, so the computer will smile upon me?''~\aff{United States~01}
\end{listquote}

A few participants also highlighted that explanations can advance human knowledge, for example, expressing concern medical progress might be slowed in the stroke diagnosis scenario if decision-making was given over to deep learning and explanations were not available.

\begin{listquote}
\item ``Doctors can perhaps even learn from it, that this has made the decision based on these grounds. Doctors might learn something, whether it's right or wrong.''~\aff{Finland~08}
\item ``{[Deep learning means]} man does not learn anything anymore... the knowledge is transferred into a system, where you cannot get it out.''~\aff{Germany~B11}
\item ``For me, the limit is, where it is better than a doctor. If I suppose that if normal doctors have a failure diagnosis rate of 25\%, then I have a higher chance to survive if a system with the 85\% probability executes the examination. The only thing that would bother me is, if we don't get any explanations and only have the system, that the knowledge about the disease and the illness will get lost.''~\aff{Germany~05}
\end{listquote}

Others spoke of explanations as building good will or demonstrating respect, consistent with discussion in the legal literature that explanations serve a dignitary and humane purpose~\cite{edwards2017slave,kaminski2019binary,selbst2018intuitive}. Some participants also wanted explanations due to broad personal preference.

\begin{listquote}
\item ``My issue with deep learning is that I've always been the type of person -- I'm not going to just take something for what it is. I need to know. I'm 100\% okay with whatever result it is, but just take me through the process of how we got to that.''~\aff{United States~07}
\item ``I'm a person that's very analytical, so I need to be able to understand, to see what it is. Let me see it, let me read it, let me understand it, let me digest it.''~\aff{United States~02}
\end{listquote}

However, some participants also argued that there were situations in which explanations were simply not useful or necessary, or would be substantially less important than other qualities such as accuracy or appeal, as shown in Table~\ref{tab:purposes}. Some participants even challenged others who said they prioritized explanations over accuracy in given scenarios, arguing that those explanations would not be helpful to them.

\begin{listquote}
\item ``If I know that it works well in 19 out of 20 cases, then I don't necessarily need to know how exactly it works.''~\aff{Germany~A12}
\item ``At the end of the day, there are things that no matter how much someone explains to me, I'm not going to understand regardless, because I know sometimes I have really intelligent friends who try to explain things to me. I'm not going to get it... I'm not going to understand everything, so if it's just going to tell me the score and then I still have the opportunity to appeal it and discuss it with someone, I'm going to go with C, because it's the most accurate.''~\aff{United States~07}
\end{listquote}

Participants often viewed straightforward decisions as not requiring explanation or human review.

\begin{listquote}
\item ``It's situational. You run a red light, it recognizes your plate. You should get a ticket for that. You don't need [an explanation]... the factors are undisputable, right?''~\aff{United States~08}
\item ``It depends on the example. If we look, for example, at a case of parking in a wrong spot and the fact that the owner of a vehicle needs to pay a fine for parking in a prohibited zone, and furthermore assume that the photo that has been taken is 100 percent accurate, I won't need a {[human]} judge. It is quite clear that my vehicle was parked in that spot.''~\aff{Germany~A03}
\item ``Maybe in areas of sports. For example, when a decision has to be made whether the person was in front or behind a line. In this case, I would rely on a technique that has a high hit-ratio [and not need additional explanation].''~\aff{Germany~A01}
\end{listquote}

In Table~\ref{tab:purposes} we summarize the ideas described in this subsection, i.e. the purposes participants thought explanations could serve as well as the reasons they sometimes saw them as unnecessary. Legal, ethical, and technical scholarship has explored some of these ideas conceptually, but to our knowledge, at best limited evidence has been provided that they resonate with members of the public. Here we draw together these disparate ideas from multiple disciplines, and provide new and supporting evidence that these ideas have a meaningful role in public discourse and policy.

\subsection{Participants felt some decisions require a human touch}

In previous sections we focused on how participants perceived algorithmic decision-makers, and how explanations might address potential issues with them. We now turn our attention to how participants perceived human decision-makers as a potential alternative or remedy to deficits in algorithmic decision-making. Participants frequently brought up humans and emphasized that humans have important decision-making capabilities that algorithmic decision-making systems do not. Relative to algorithmic decision-making systems, humans were seen (in some or all situations) as more trustworthy, compassionate and capable of treating people well, capable of complex judgment and assessing intangible factors, and engaging in negotiation. Decision-making tasks involving such capabilities were seen as best left to humans. Further, participants viewed humans as having a critical role in overseeing and intervening in algorithmic decision-making systems, in order to prevent the loss of autonomy discussed in Section~\ref{sec:mysterious}.

Participants often brought up the importance of `the human touch' or `the human element' in decision-making. Human knowledge, collective or individual, was often seen as more trustworthy than machine knowledge or artificial intelligence. Some participants observed that System A `had the most human element' because it was based on human thinking and the knowledge of human professionals, and preferred it for that reason.

\begin{listquote}
\item ``I somehow wish for a human for this kind of thing {[stroke diagnosis]}.''~\aff{Finland~03}
\item ``I am worried about moving to a future where there's literally no human element. I really am.''~\aff{United States~09}
\item ``I think there's that element of us as humans, just not being able to trust anything that's not us.''~\aff{United Kingdom~08}
\end{listquote}

As mentioned in Section~\ref{sec:technical}, AI was characterized as lacking human capabilities such as emotion, empathy, or compassion. Relatedly, the lack of human involvement in a decision was sometimes characterized as devaluing the recipient. Human participation was viewed as adding emotional capacity and respect to the decision-making process.

\begin{listquote}
\item ``That's the human factor, someone who explains it empathically, not just the machine, which says `You're bad'.''~\aff{Finland~10}
\item ``I mean, the robot can't be merely a robot. It has to sort of adopt some human features of empathy and compassion.''~\aff{United~Kingdom~09}
\end{listquote}

As also discussed in Section~\ref{sec:technical}, AI was typically seen as lacking the capacity to make complex moral judgments or assess soft factors. By contrast, humans were viewed as strong at these tasks. Therefore, participants felt decision-making responsibilities involving these more subjective judgments should be given to humans rather than AI.

Human involvement was called out as particularly important for negotiation. Some participants highlighted the importance of going beyond formal decision-making criteria, and having flexibility to work around the system. This moved beyond assessment of intangibles like soft factors, to actions like considering extenuating circumstances, granting leniency for catastrophic events in people's lives, `giving people a chance,' or taking into account personal trust.

\begin{listquote}
\item ``I prefer a human-based one because I can say, `Look, I've got some of the criteria that you've asked for, but the others, I'm willing to learn. I'm a fast learner. Hire me, for a month, and if I haven't learnt it, then obviously get someone else in. But I'm prepared to learn. And I'll try my best.'{}''~\aff{United~Kingdom~02}
\item ``I work in social housing, and there's some questions that are asked all the time. And rather than having someone answer the same question all the time [chuckling] ... do you know what I mean? `How do I pay my rent?' Those basic questions. A chatbot can do that. But if it's, `I can't pay my rent because ...' `I can pay universal credit' or something else, for me, a human will always have to do that.''~\aff{United~Kingdom~12}
\item ``I mean, if you went in the Old West, into a bank and you're sitting down with the manager and they say, `Well, I trust you, I know your family, let me give you a loan.' That's the human element. There is no human element in the credit system.''~\aff{United~States~08}
\end{listquote}

Humans were also seen as more variable than algorithmic decision-making systems, with approaches to multiple humans offering another avenue to get a favorable decision. Algorithmic decision-making systems were seen as consistent within institutions, and perhaps even across institutions.

\begin{listquote}
\item \affpre{United States 01}: ``The human ability to make judgment calls. We are not all the same, with the same parameters. There are times within the human experience --''\newline
\affpre{United States 02}: ``Things happen.''\newline
\affpre{United States 01}: ``Things happen. I mentioned, your kid falls out of a tree, you all of a sudden have huge medical bills, which changes your whole ability to handle other finances in your life and this is why I think appeal is important. And here’s another thing, it depends on who you get on the phone.''\newline
...\newline
\affpre{United States 02}: ``That's true, because I've gotten people on the phone when I spoke to my credit card company -- like I'll speak to somebody at 11:00 and I may speak to somebody at 1:00 --''\newline
\affpre{United States 01}: ``You get two different --''\newline
\affpre{United States 02}: ``You get two different answers. Two different temperaments.''
\end{listquote}

Finally, humans were seen as having an essential role in decision-making systems. 
Human oversight was viewed as necessary to keep control over machines and ensure they were behaving in a socially responsible manner, for example, by periodically auditing the system. Further, it was seen as advantageous to integrate humans in particular circumstances or at moments when algorithmic decision-making might be flawed, for example to use a human instead or have a human review output from the algorithm to make a final decision.
However, other research points to challenges in successfully allocating responsibilities between humans and AI, as we discuss further in Section~\ref{sec:hitl}.

\subsection{Comparison of Human and AI Decision-Making}

\input{figtab/qualities}

In Table~\ref{tab:qualities} we summarize the desirable qualities of good decision-makers, whether human or AI, that were raised during the workshops.\footnote{While some of the items on the list were presumably influenced by information provided during the workshops, workshop discussion was open-ended. Participants sometimes disagreed with or challenged the information provided, and often raised new issues that had not been discussed. A follow-up study, either qualitative or quantitative, could be performed to determine the extent to which participants generate these concepts spontaneously. See also the related discussion of advantages and disadvantages of automated decision-making in~\cite{singh2019} which mentions a subset of these issues.} We note that while consistency is often positioned as an advantage of automation, as discussed above, some participants saw downsides in removing human variability and variability across institutions, as that variability gave them more `chances' to get a favorable decision.

%
%


Although there is some debate on this topic, the predominant position appears to be that all these qualities cannot be simultaneously achieved in entirety, leading to the need to prioritize. We now discuss the relative value of these qualities to participants when considering: (1) which factors were most valued for an algorithmic decision-making system, in a given situation; and (2) which factors would prompt more or less human versus AI involvement, in a given situation. Considering tradeoffs between different algorithmic decision-making systems, qualities such as accuracy, speed, scale, explanation, and challenge were seen as critically important for at least one of the scenarios presented, and often more. For example, participants often emphasized accuracy and speed in the stroke diagnosis scenario.

\begin{listquote}
\item ``You go for accuracy when your life is at stake.''~\aff{United States~03}
\item ``In this kind of stroke, the faster the correct decision comes, the better it is... There's no time for nitpicking.''~\aff{Finland~11}
\item \affpre{Finland 02}: ``If I were the patient, I would be interested especially in accuracy. If we think about the 25\% probability of getting a wrong diagnosis, I'm more interested in whether the diagnosis is correct, instead of the explanation of the diagnosis, in this case...''\newline
\affpre{Finland 08}: ``Yes. It doesn't give much consolation to get a good explanation if you're already in a casket.''
\end{listquote}

A small number of participants called out a potential downside of accuracy, that decisions could be harder to overturn if the algorithmic decision-making system were known to be highly accurate. Therefore, they expressed a preference for lower accuracy in some situations because they thought it would be easier to successfully appeal if the system were less trusted, resonating with the concept of ambiguity as a design resource~\cite{aoki2005,gaver2003}.

\begin{listquote}
\item  ``One of the reasons I chose A is because I thought that if it was going to be that inaccurate -- only 75\% -- it wouldn't be trusted as much, like it would maybe be used to speed up the investigations or something, but it wouldn't be used as evidence.''~\aff{United States 10}
\end{listquote}

Participants sometimes preferred a system that balances competing priorities.

\begin{listquote}
\item ``{[System B is]} a decent compromise. I'm willing to give up some accuracy to get more communication, more of a two-way street of dialogue, versus the machine saying, `This is my final roll in, lots of luck.' ... So, it's a good middle ground.''~\aff{United States~01}
\item ``I am the kind of person who takes the middle road. There is still room for a little bit of freedom. You can influence the system and you can trace the process. It is not the extreme system of the three of them.''~\aff{Germany~A11}
\item ``{[System B is]} the middle path. Fairly precise, but gives the possibility of challenging it.''~\aff{Finland~01}
\end{listquote}

Situational factors influenced not only preference among different algorithmic decision-making systems (e.g. one that was more accurate but provided less explanation, or vice versa), but also perception of whether a human must be involved in the decision-making process, for example, providing oversight over an algorithmic system and making the final decision, or whether an algorithmic decision-making system was seen as appropriate to implement at all versus relying solely on a human decision-maker.
 
\begin{listquote}
\item ``It depends so much on the situation. There are situations where {[AI]} could be and surely also is more precise than a human.''~\aff{Finland~08}
\item ``I think that it really depends on the topic. If we are talking about face recognition and stroke, you are talking about health and your civil rights and whether you'll be detained or not, so it's about your freedom. Those are two absolutely extreme things concerning you and I would never give any AI the full and 100\% authority to make a decision.''~\aff{Germany~B11}
\item ``I've experienced something like that in my family a year ago. A life-saving surgery had to happen, and parts of it were done by a robot, which couldn't be done by humans. And my dad wouldn't be alive if that wouldn't have happened... so I thought, okay there isn't any doubting anymore, you just try it, because it is the only thing that can help. And then, that is the thing you cling to. In that moment, completely other things matter. When you know there is no other way, then you do it. It is maybe also in relation to what the initial circumstances are, are there still many options, or aren't there any left? And then it is maybe the only possibility. And in this case, I believe it is good that it exists, and that the development got so far, that they are more accurate and more unerring and can do things that humans aren't able to.''~\aff{Germany~B09}
\end{listquote}

Overall, considering tradeoffs such as accuracy versus explainability, a number of situational factors were important. Some of the most influential factors were: whether the decision was mechanical versus subjective; the severity of the consequences; whether it was the recipient's only chance at a decision (e.g. a medical decision or legal trial might offer only one meaningful opportunity at a decision, whereas an individual could apply multiple times for a loan); whether the recipient could take meaningful action based on having information; or the perspective of different parties (for example, for an algorithmic decision-making system that screens job applications, an applicant might prioritize explanations but a hiring manager might prioritize accuracy).

\begin{listquote}
\item ``It varies between scenarios... it just depends on the situation.''~\aff{United States~07}
\item ``This is obviously different from the first one {[job application screening]}, where this {[credit scoring]} is very black and white. You either did or you did not pay on time your mortgage bill, your credit card, blah, blah, blah. And so initially you think, this is so accurate you go with C. However, this is so much more important than a job interview, because if you get an inaccurate credit rejection and you can't find out why, that's really concerning. Because that... will affect your life.''~\aff{United Kingdom~11}
\item ``Facial recognition affects us on a much more personal level.''~\aff{Germany~A11}
\item ``The effects of the inaccuracy are more damaging {[for law enforcement]} than with credit. I mean, a wrong facial recognition could change a life.''~\aff{United States~05}
\item ``If you are all of a sudden part of the 25\% and you have been categorised in the wrong way and then you are up for a crime... It goes much deeper than a decision about whether an application is accepted or not.''~\aff{Germany~A02}
\item ``System A becomes more relevant in terms of the consequences that result from it. If there is a threat to go to jail, it is more important for me that the system is transparent and that I can prove it was not me. System A might be less relevant in cases where I apply for a job and am declined, because I can apply to other companies.''~\aff{Germany~A06}
\item ``I would like to know how it functions, if it makes decisions this big, as medical ones or legal ones, in this country. It's important that there's access to see how it functions.''~\aff{Finland~11}
\item ``From the company's point of view, it would be the accuracy. But from someone applying, it would be the explanation {[murmurs of agreement]}.''~\aff{United Kingdom~10}
\end{listquote}

%

%% file: figtab/purposes.tex
\begin{table}
\begin{tabular}{ p{8.1cm}p{6.2cm} }
\fontsize{8.5pt}{8.1pt}\selectfont

\textbf{Purposes of Explanations} & \textit{Notable Mentions in Other Work} \\
\arrayrulecolor{black}\midrule
Enable individual challenge & Legal~\cite{edwards2017slave, selbst2018intuitive, kaminski2019right, kaminski2019binary, citron2014, wachter2018counterfactual}; \\
 & Deliberative Democracy~\cite{ico2019}; \\
 & Industry Roundtable~\cite{ico2019};
  Technical~\cite{tintarev2007} \\   \arrayrulecolor{ltgray}\midrule
Inform personal behavior
 & Legal~\cite{selbst2018intuitive, wachter2018counterfactual};
   Deliberative Democracy~\cite{ico2019,oswald2019}; \\
 & Industry Roundtable~\cite{ico2019} \\ \midrule
Enable accountability \& oversight
 & Legal~\cite{selbst2018intuitive,kaminski2019binary,citron2014}; 
   Deliberative Democracy~\cite{ico2019} \\ \midrule
Be respectful & Legal ~\cite{edwards2017slave,selbst2018intuitive,kaminski2019binary} \\ \midrule
Build user trust \& satisfaction & Technical ~\cite{biran2017,herman2017,tintarev2007} \\ \midrule
Advance human knowledge & -- \\ \midrule
I'm the kind of person who needs explanations & -- \\
\arrayrulecolor{black}\midrule
\\[-0.1cm]
\textbf{Reasons explanations may be seen as unnecessary} & \\
\midrule
May not understand explanation & Deliberative Democracy~\cite{ico2019} \\ \arrayrulecolor{ltgray}\midrule
Bad time to receive explanation & Deliberative Democracy~\cite{ico2019} \\ \midrule
Trust system (e.g. system accuracy has been verified) & Ethics~\cite{london2019}; Technical~\cite{doshi2017} \\ \midrule
Don't require an explanation from humans in that situation & Deliberative Democracy~\cite{oswald2019} \\  \midrule
Reason for decision is obvious & -- \\ \midrule
Already know reason for decision & -- \\
\arrayrulecolor{black}\bottomrule
\end{tabular}
\caption{In the lefthand column we first list purposes participants discussed in our study and then reasons why explanations were seen by our participants as unnecessary, as described in Section~\ref{sec:explanations}. In the righthand column we draw connections with notable mentions of these ideas in legal, technical and other literatures. Previous work has been largely conceptual, with rare mentions in user research. Our work provides new and supporting evidence that these ideas resonate with members of the public.}
\label{tab:purposes}
\end{table}

%% file: figtab/qualities.tex
\begin{table}
\begin{tabular}{ l l l }

  & \multicolumn{2}{l}{\textit{Perceived to do well}} \\
  & \textbf{Humans} & \textbf{AI} \\
\midrule
Speed & No & Yes \\
Scale & No & Yes \\
Consistency & No & Yes \\
Objectivity & No & Mixed \\
Accuracy & Mixed & Mixed \\
Provide Explanations & Mixed & Mixed \\
Provide Opportunity to Challenge & Mixed & Mixed \\
Provide Opportunity to Negotiate & Yes & No \\
Provide Human Touch & Yes & No \\
Make Subjective Judgments & Yes & No \\
Make Straightforward Decisions & Yes & Yes \\
\bottomrule
\end{tabular}
\caption{Desirable qualities of decision-makers.}
\label{tab:qualities}
\end{table}

%% file: 09discussion.tex
Our findings suggest productive directions for design, technical and research approaches to algorithmic decision-making.

\subsection{Design choices related to explainability should account for situational variation}
Our research demonstrated that tradeoffs change across scenarios, for example, in some cases explanation, appeal, or other forms of transparency were a higher priority than accuracy but in other cases the reverse was true. Our research therefore suggests that compromising other qualities of an algorithmic decision-making system in order to provide an explanation is not always the right choice. We also highlight that while severity was sometimes a factor, it was often not the dominant consideration and was not sufficient to predict whether participants felt an explanation should be required. These findings reinforce and extend previous arguments that broad requirements to provide explanations are unlikely to meet the full spectrum of public needs~\cite{balaram2018, ico2019, oswald2019, singh2019}. Our research also underscores that explanation per se is not necessarily the end goal~\cite{kluttz2019,rader2018,woodruff2019}, but rather there are goals underlying requests for explanation that can often be met via other mechanisms that provide accountability, contestability, appeal, advancement of human knowledge, or other qualities.

\subsection{AI's (perceived) incapacity for humanity and subjectivity}
AI was seen as lacking human qualities, such as the ability to show empathy, or the ability to make subjective decisions that involved complex moral judgments or intangible qualities. These perceptions are likely accurate for most of the algorithmic decision-making systems people have interacted with and that are in production today. However, future systems may be designed with more of these qualities as goals. Further, even for today's systems, perception of AI as highly inflexible may not be entirely accurate. Communication and education about AI might be valuable towards influencing the public's perception of its qualities.

\subsection{Negotiability in decision-making systems}
Our findings surfaced a new concept that is on the periphery of related capabilities in the literature such as contestability and recourse~\cite{hirsch2017,kluttz2019,ustun2019,vaccaro2019,venkatasubramanian2020}.
Discussion of these capabilities generally focuses on a decision-making framework, for example, examining how to challenge decisions that have not been made correctly within that framework, how experts may engage with the system to modify the framework, or how individuals can receive guidance on how they can change their behavior to meet the criteria of the framework. We propose negotiability, the ability of a decision-making system to allow individuals to request that a decision be modified not because it is based on inaccurate data and not because the decision is incorrect, but rather because they would like to ask the decision-maker to go beyond the established decision-making framework and criteria, to take into account factors outside of the accepted decision-making process, such as extenuating circumstances or personal relationships. In existing encounters with institutional decision-making, participants had sometimes been able to navigate human variability, empathy, and inconsistency to get a favorable result, and they felt that this opportunity would be lost with a non-human decision-maker. This connected with their beliefs that algorithmic decision-making systems were dispassionate, authoritarian, inflexible, and consistent, and would not be susceptible to appeals for compassion or special treatment. Such negotiability is not without challenges and drawbacks. Institutional movement towards automation and bureaucracy in many ways seeks to eliminate exactly such inconsistency and attendant concerns such as unfairness or bias. 
Even with these limitations, negotiability is a feature that some respondents appear to value in existing approaches, and it is worth considering further whether or how this need might be met (whether by algorithmic systems or human review) while also preserving other important qualities.

\subsection{Further research on human-in-the-loop}
\label{sec:hitl}
Participants saw strong advantages to AI in many cases, and participant interest in instilling humanity in decision-making processes raises interesting questions about the role of humans-in-the-loop~\cite{dearteaga2020,fjeld2020}.
Participants were highly interested in the roles humans and AI would play at different moments in the decision-making process, and their opinions regarding whether the use of AI was beneficial for a given use case often rested on detailed specifics of roles played by humans versus AI in the overall decision-making process.
Further, well-executed human-in-the-loop strategies have the potential to ameliorate concerns about autonomy.
Given that participants perceived that AI and humans had many complementary strengths,
it would be productive to further explore how to effectively allocate responsibility 
and tasks between them in algorithmic decision-making systems.
At the same time, other research has shown that handoffs between humans and AI are complex and difficult to design well~\cite{elish2019, green2019}.
We believe this to be an underexplored area and highlight this as a valuable topic for future research.

%% file: 11conclusion.tex
We reported on a qualitative study of perceptions of algorithmic decision- making, conducted in Finland, Germany, the United Kingdom, and the United States. Participants shared thoughtful, nuanced insights on the qualities of human and AI decision-makers. Participants saw strong advantages to AI in many circumstances, but in some cases also felt that at certain moments humans should assume important responsibilities in algorithmic decision-making systems. We found high situational variance in the prioritization of different qualities, for example, explainability and appeal were prioritized in certain situations but not others. Further, algorithmic decision-making has been heralded as offering many benefits in terms of efficiency and consistency. While participants saw some of these qualities as desirable, they saw others as undesirable, and associated them with a loss of humanity, judgment, and flexibility. Future efforts can explore whether these qualities are inherently in tension with a shift towards algorithmic decision-making, and whether AI can effectively support perceived humanity and negotiability.

%% file: figtab/methodtab.tex
\begin{table}[ht]
      \centering
      \begin{tabular}{p{3.2cm} p{11.3cm}}
           \toprule
           Scenario & Description \\ \midrule
           \fontsize{8.5pt}{8.1pt}\selectfont
           \textit{Stroke Diagnosis} & 
              Rapid and accurate diagnosis of stroke greatly increases chances of survival and recovery of the patient. This is highly specialised work which ideally should be done by neuroradiologists with many years of training and experience. However, these experts are not available in every hospital, 24 hours a day, 7 days a week, and in practice diagnosis is often done by nonspecialist emergency medicine doctors. As diagnostic data are accumulated from previous stroke patients, AI systems could provide stroke diagnosis that is fast, and always available in hospitals. 
              \\[2.4cm] & 
              \textit{Draws heavily from the NIHR Greater Manchester Patient Safety Translational Research Centre and the Information Commissioner’s Office (ICO) Citizens' Juries work \url{http://www.bit.ly/GMPSTRCCitizensJuries}} \\ \arrayrulecolor{ltgray}\midrule
           \textit{Job Applicant Screening} & 
              A large company in the private sector receives a high volume of applications, and plans to use an AI system to screen job applications and make shortlisting decisions. This will allow the company to screen every application and process applications more quickly. This system may also more objectively make decisions that would otherwise be prone to human errors or judgment. On the other hand, depending on how the system is trained, it may learn to reinforce existing patterns of hiring. 
              \\[2.0cm] & \textit{Draws heavily on the ICO and RSA (Royal Society for the encouragement of Arts, Manufactures and Commerce) Citizens' Juries work~\cite{balaram2018}}\\ \midrule
           \textit{Credit Scoring} & 
              A new credit scoring company has decided to enter the market, and provide a new credit score. They would like their scores to be more accurate, and are considering using AI to assign people scores in the hopes of better predicting whether they will repay money. Using AI will allow for many more features to be input and also pick out a more accurate way to combine features to predict a person’s likelihood of repaying. AI advocates and experts think this will allow companies to lend more money to people who are likely to pay it back, but who might have been excluded in the traditional system. 
              \\[2.4cm] & \textit{New, but fits within the generic ICO format}\\ \midrule
           \textit{Law Enforcement} &  
              Imagine that local law enforcement in your city is under-resourced, sometimes causing it to fail to prevent crimes or leave crimes unsolved. The city government decides to procure a facial recognition system to make law enforcement more efficient. The system will allow law enforcement to identify people or verify their identities, for example, to find missing persons, match surveillance footage from a crime scene against a database of faces, or verify people’s identities when they are stopped by the police. 
              \\[2.0cm] & \textit{New, but fits within the generic ICO format}\\
           \arrayrulecolor{black}\bottomrule
      \end{tabular}
      \caption{Excerpts from each of the four hypothetical scenarios.}
      \label{tab:scenarios}
  \end{table}


\begin{table}[ht]
      \centering
      \begin{tabular}{ l p{4.2cm} p{4.2cm} p{3.5cm} }
           \toprule
            & System A & System B & System C \\[-0.1em]
            & \textbf{\small Rule-Based} 
              & \textbf{\small Conventional Machine Learning} 
              & \textbf{\small Deep Learning} \\[0.4em]

  \multicolumn{2}{l}{\normalsize \emph{Stroke Diagnosis}} \\ \midrule        
  \textbf{Accuracy} & 
     75\% (nonspecialist doctor's level) &
     85\% (expert doctor's level) &
     95\% (beyond human level) \\ \arrayrulecolor{ltgray}\midrule 
     
  \textbf{Transparency} & 
      Full explanation of all the specific reasons that led to diagnosis &
      Partial explanation describing features that played a role in diagnosis &
      No explanation \\ \arrayrulecolor{black}\midrule  \\[-0.2em]

  \multicolumn{2}{l}{\normalsize \emph{Law Enforcement}} \\ \midrule
  \textbf{Accuracy} & 
      75\% (accurate in good lighting and conditions) &
      85\% (highly accurate) &
      95\% (most accurate AI technology) \\ \arrayrulecolor{ltgray}\midrule

  \textbf{Transparency} &
       Provides an ability for people who have had their faces matched to appeal (request human review prior to formal decision-making), plus monthly reports for regular external oversight &
       Provides an ability for people who have had their faces matched to appeal (request human review prior to formal decision-making), but no process for regular external oversight &
       No process to appeal (request human review prior to formal decision-making), no process for regular external oversight \\[0.2em]
      \arrayrulecolor{black}\bottomrule
      \end{tabular}
      \caption{Descriptions of the three systems participants could choose among for the Stroke Diagnosis and Law Enforcement scenarios. For all scenarios, participants were offered three systems with 75\%, 85\%, and 95\% accuracy, where the most accurate system provided the least transparency. In all scenarios except Law Enforcement, transparency was accomplished through explanations; for Law Enforcement we offered transparency options focused on appeal and oversight.}
      \label{tab:systems}
  \end{table}

%% file: 00manuscript.bbl

\begin{thebibliography}{101}


\ifx \showCODEN    \undefined \def \showCODEN     #1{\unskip}     \fi
\ifx \showDOI      \undefined \def \showDOI       #1{#1}\fi
\ifx \showISBNx    \undefined \def \showISBNx     #1{\unskip}     \fi
\ifx \showISBNxiii \undefined \def \showISBNxiii  #1{\unskip}     \fi
\ifx \showISSN     \undefined \def \showISSN      #1{\unskip}     \fi
\ifx \showLCCN     \undefined \def \showLCCN      #1{\unskip}     \fi
\ifx \shownote     \undefined \def \shownote      #1{#1}          \fi
\ifx \showarticletitle \undefined \def \showarticletitle #1{#1}   \fi
\ifx \showURL      \undefined \def \showURL       {\relax}        \fi
\providecommand\bibfield[2]{#2}
\providecommand\bibinfo[2]{#2}
\providecommand\natexlab[1]{#1}
\providecommand\showeprint[2][]{arXiv:#2}

\bibitem[\protect\citeauthoryear{Abdul, Vermeulen, Wang, Lim, and
  Kankanhalli}{Abdul et~al\mbox{.}}{2018}]%
        {abdul2018}
\bibfield{author}{\bibinfo{person}{Ashraf Abdul}, \bibinfo{person}{Jo
  Vermeulen}, \bibinfo{person}{Danding Wang}, \bibinfo{person}{Brian~Y. Lim},
  {and} \bibinfo{person}{Mohan Kankanhalli}.} \bibinfo{year}{2018}\natexlab{}.
\newblock \showarticletitle{Trends and trajectories for explainable,
  accountable and intelligible systems: An HCI research agenda}. In
  \bibinfo{booktitle}{\emph{Proceedings of the 2018 CHI Conference on Human
  Factors in Computing Systems}}. \bibinfo{pages}{1--18}.
\newblock
\urldef\tempurl%
\url{https://doi.org/10.1145/3173574.3174156}
\showDOI{\tempurl}


\bibitem[\protect\citeauthoryear{Ananny and Crawford}{Ananny and
  Crawford}{2016}]%
        {ananny2016}
\bibfield{author}{\bibinfo{person}{Mike Ananny} {and} \bibinfo{person}{Kate
  Crawford}.} \bibinfo{year}{2016}\natexlab{}.
\newblock \showarticletitle{Seeing without knowing: Limitations of the
  transparency ideal and its application to algorithmic accountability}.
\newblock \bibinfo{journal}{\emph{New Media \& Society}}
  (\bibinfo{year}{2016}).
\newblock
\urldef\tempurl%
\url{https://doi.org/10.1177/1461444816676645}
\showDOI{\tempurl}


\bibitem[\protect\citeauthoryear{Aoki and Woodruff}{Aoki and Woodruff}{2005}]%
        {aoki2005}
\bibfield{author}{\bibinfo{person}{Paul~M. Aoki} {and} \bibinfo{person}{Allison
  Woodruff}.} \bibinfo{year}{2005}\natexlab{}.
\newblock \showarticletitle{Making space for stories: ambiguity in the design
  of personal communication systems}. In \bibinfo{booktitle}{\emph{Proceedings
  of the SIGCHI Conference on Human Factors in Computing Systems (CHI 2005)}}.
  \bibinfo{pages}{181--190}.
\newblock
\urldef\tempurl%
\url{https://doi.org/10.1145/1054972.1054998}
\showDOI{\tempurl}


\bibitem[\protect\citeauthoryear{Armour}{Armour}{1995}]%
        {armour1995}
\bibfield{author}{\bibinfo{person}{Audrey Armour}.}
  \bibinfo{year}{1995}\natexlab{}.
\newblock \showarticletitle{The citizens' jury model of public participation: a
  critical evaluation}.
\newblock In \bibinfo{booktitle}{\emph{Fairness and Competence in Citizen
  Participation}}. \bibinfo{publisher}{Springer}, \bibinfo{pages}{175--187}.
\newblock


\bibitem[\protect\citeauthoryear{Balaram, Greenham, and Leonard}{Balaram
  et~al\mbox{.}}{2018}]%
        {balaram2018}
\bibfield{author}{\bibinfo{person}{Brhmie Balaram}, \bibinfo{person}{Tony
  Greenham}, {and} \bibinfo{person}{Jasmine Leonard}.}
  \bibinfo{year}{2018}\natexlab{}.
\newblock \bibinfo{title}{Artificial Intelligence: Real Public Engagement}.
\newblock
\newblock
\urldef\tempurl%
\url{www.thersa.org/discover/publications-and-articles/reports/artificial-intelligence-real-public-engagement}
\showURL{%
\tempurl}


\bibitem[\protect\citeauthoryear{Balog and Radlinski}{Balog and
  Radlinski}{2020}]%
        {balog2020}
\bibfield{author}{\bibinfo{person}{Krisztian Balog} {and}
  \bibinfo{person}{Filip Radlinski}.} \bibinfo{year}{2020}\natexlab{}.
\newblock \showarticletitle{Measuring Recommendation Explanation Quality: The
  Conflicting Goals of Explanations}. In \bibinfo{booktitle}{\emph{Proceedings
  of the 43rd International ACM SIGIR Conference on Research and Development in
  Information Retrieval}}. \bibinfo{pages}{329–338}.
\newblock
\urldef\tempurl%
\url{https://doi.org/10.1145/3397271.3401032}
\showDOI{\tempurl}


\bibitem[\protect\citeauthoryear{Barocas and Selbst}{Barocas and
  Selbst}{2016}]%
        {barocas2016}
\bibfield{author}{\bibinfo{person}{Solon Barocas} {and}
  \bibinfo{person}{Andrew~D. Selbst}.} \bibinfo{year}{2016}\natexlab{}.
\newblock \showarticletitle{Big Data's Disparate Impact}.
\newblock \bibinfo{journal}{\emph{California Law Review}}
  \bibinfo{volume}{104} (\bibinfo{year}{2016}), \bibinfo{pages}{671}.
\newblock


\bibitem[\protect\citeauthoryear{Beyer and Holtzblatt}{Beyer and
  Holtzblatt}{1997}]%
        {beyer1997contextual}
\bibfield{author}{\bibinfo{person}{Hugh Beyer} {and} \bibinfo{person}{Karen
  Holtzblatt}.} \bibinfo{year}{1997}\natexlab{}.
\newblock \bibinfo{booktitle}{\emph{Contextual Design: Defining
  Customer-centered Systems}}.
\newblock \bibinfo{publisher}{Elsevier}.
\newblock


\bibitem[\protect\citeauthoryear{Biran and Cotton}{Biran and Cotton}{2017}]%
        {biran2017}
\bibfield{author}{\bibinfo{person}{Or Biran} {and} \bibinfo{person}{Courtenay
  Cotton}.} \bibinfo{year}{2017}\natexlab{}.
\newblock \showarticletitle{Explanation and justification in machine learning:
  A survey}. In \bibinfo{booktitle}{\emph{IJCAI 2017 Workshop on Explainable
  Artificial Intelligence (XAI)}}, Vol.~\bibinfo{volume}{8}.
  \bibinfo{pages}{8--13}.
\newblock


\bibitem[\protect\citeauthoryear{boyd and Crawford}{boyd and Crawford}{2012}]%
        {boyd2012}
\bibfield{author}{\bibinfo{person}{danah boyd} {and} \bibinfo{person}{Kate
  Crawford}.} \bibinfo{year}{2012}\natexlab{}.
\newblock \showarticletitle{Critical Questions For Big Data}.
\newblock \bibinfo{journal}{\emph{Information Communication \& Society}}
  \bibinfo{volume}{15}, \bibinfo{number}{5} (\bibinfo{year}{2012}),
  \bibinfo{pages}{662--679}.
\newblock
\urldef\tempurl%
\url{https://doi.org/10.1080/1369118X.2012.678878}
\showDOI{\tempurl}


\bibitem[\protect\citeauthoryear{Breiman et~al\mbox{.}}{Breiman
  et~al\mbox{.}}{2001}]%
        {breiman2001}
\bibfield{author}{\bibinfo{person}{Leo Breiman} {et~al\mbox{.}}}
  \bibinfo{year}{2001}\natexlab{}.
\newblock \showarticletitle{Statistical modeling: The two cultures (with
  comments and a rejoinder by the author)}.
\newblock \bibinfo{journal}{\emph{Statist. Sci.}} \bibinfo{volume}{16},
  \bibinfo{number}{3} (\bibinfo{year}{2001}), \bibinfo{pages}{199--231}.
\newblock


\bibitem[\protect\citeauthoryear{Bucher}{Bucher}{2017}]%
        {bucher2017algorithmic}
\bibfield{author}{\bibinfo{person}{Taina Bucher}.}
  \bibinfo{year}{2017}\natexlab{}.
\newblock \showarticletitle{The algorithmic imaginary: exploring the ordinary
  affects of Facebook algorithms}.
\newblock \bibinfo{journal}{\emph{Information, Communication \& Society}}
  \bibinfo{volume}{20}, \bibinfo{number}{1} (\bibinfo{year}{2017}),
  \bibinfo{pages}{30--44}.
\newblock


\bibitem[\protect\citeauthoryear{Burrell}{Burrell}{2016}]%
        {burrell2016machine}
\bibfield{author}{\bibinfo{person}{Jenna Burrell}.}
  \bibinfo{year}{2016}\natexlab{}.
\newblock \showarticletitle{How the machine ‘thinks’: Understanding opacity
  in machine learning algorithms}.
\newblock \bibinfo{journal}{\emph{Big Data \& Society}} \bibinfo{volume}{3},
  \bibinfo{number}{1} (\bibinfo{year}{2016}).
\newblock
\urldef\tempurl%
\url{https://doi.org/10.1177/2053951715622512}
\showDOI{\tempurl}


\bibitem[\protect\citeauthoryear{Cai, Winter, Steiner, Wilcox, and Terry}{Cai
  et~al\mbox{.}}{2019}]%
        {cai2019}
\bibfield{author}{\bibinfo{person}{Carrie~Jun Cai}, \bibinfo{person}{Samantha
  Winter}, \bibinfo{person}{David Steiner}, \bibinfo{person}{Lauren Wilcox},
  {and} \bibinfo{person}{Michael Terry}.} \bibinfo{year}{2019}\natexlab{}.
\newblock \showarticletitle{``Hello AI": Uncovering the Onboarding Needs of
  Medical Practitioners for Human-AI Collaborative Decision-Making}. In
  \bibinfo{booktitle}{\emph{Proceedings of the 2019 ACM Conference on Computer
  Supported Cooperative Work and Social Computing (CSCW '19)}}.
\newblock
\urldef\tempurl%
\url{https://doi.org/10.1145/3359206}
\showDOI{\tempurl}


\bibitem[\protect\citeauthoryear{Capital}{Capital}{2019}]%
        {blumberg2019}
\bibfield{author}{\bibinfo{person}{Blumberg Capital}.}
  \bibinfo{year}{2019}\natexlab{}.
\newblock \bibinfo{title}{Artificial Intelligence in 2019: Getting Past the
  Adoption Tipping Point}.
\newblock
\newblock


\bibitem[\protect\citeauthoryear{Caruana, Simard, Weinberger, and
  LeCun}{Caruana et~al\mbox{.}}{2017}]%
        {caruana2017}
\bibfield{author}{\bibinfo{person}{Rich Caruana}, \bibinfo{person}{Patrice
  Simard}, \bibinfo{person}{Kilian Weinberger}, {and} \bibinfo{person}{Yann
  LeCun}.} \bibinfo{year}{2017}\natexlab{}.
\newblock \bibinfo{title}{The Great AI Debate -- Position: Interpretability is
  necessary for machine learning, for and against}.
\newblock \bibinfo{howpublished}{The 31st Conference on Neural Information
  Processing Systems (NIPS 2017)}.
\newblock


\bibitem[\protect\citeauthoryear{Cave, Coughlan, and Dihal}{Cave
  et~al\mbox{.}}{2019}]%
        {cave2019}
\bibfield{author}{\bibinfo{person}{Stephen Cave}, \bibinfo{person}{Kate
  Coughlan}, {and} \bibinfo{person}{Kanta Dihal}.}
  \bibinfo{year}{2019}\natexlab{}.
\newblock \showarticletitle{"Scary Robots": Examining Public Responses to AI}.
  In \bibinfo{booktitle}{\emph{Proceedings of the 2019 AAAI/ACM Conference on
  AI, Ethics, and Society (AIES 2019)}}. \bibinfo{pages}{331--337}.
\newblock
\urldef\tempurl%
\url{https://doi.org/10.1145/3306618.3314232}
\showDOI{\tempurl}


\bibitem[\protect\citeauthoryear{Cave, Craig, Dihal, Dillon, Montgomery,
  Singler, and Taylor}{Cave et~al\mbox{.}}{2018}]%
        {cave2018portrayals}
\bibfield{author}{\bibinfo{person}{Stephen Cave}, \bibinfo{person}{Claire
  Craig}, \bibinfo{person}{Kanta~Sarasvati Dihal}, \bibinfo{person}{Sarah
  Dillon}, \bibinfo{person}{Jessica Montgomery}, \bibinfo{person}{Beth
  Singler}, {and} \bibinfo{person}{Lindsay Taylor}.}
  \bibinfo{year}{2018}\natexlab{}.
\newblock \showarticletitle{Portrayals and perceptions of AI and why they
  matter}.
\newblock  (\bibinfo{year}{2018}).
\newblock


\bibitem[\protect\citeauthoryear{Center}{Center}{2004}]%
        {jefferson2004}
\bibfield{author}{\bibinfo{person}{The~Jefferson Center}.}
  \bibinfo{year}{2004}\natexlab{}.
\newblock \bibinfo{title}{The Citizens' Jury Handbook}.
\newblock
\newblock


\bibitem[\protect\citeauthoryear{Citron}{Citron}{2007}]%
        {citron2007}
\bibfield{author}{\bibinfo{person}{Danielle~Keats Citron}.}
  \bibinfo{year}{2007}\natexlab{}.
\newblock \showarticletitle{Technological Due Process}.
\newblock \bibinfo{journal}{\emph{Washington University Law Review}}
  \bibinfo{volume}{85} (\bibinfo{year}{2007}), \bibinfo{pages}{1249--1313}.
\newblock


\bibitem[\protect\citeauthoryear{Citron and Pasquale}{Citron and
  Pasquale}{2014}]%
        {citron2014}
\bibfield{author}{\bibinfo{person}{Danielle~Keats Citron} {and}
  \bibinfo{person}{Frank Pasquale}.} \bibinfo{year}{2014}\natexlab{}.
\newblock \showarticletitle{The Scored Society: Due Process for Automated
  Predictions}.
\newblock \bibinfo{journal}{\emph{Washington Law Review}}  \bibinfo{volume}{89}
  (\bibinfo{year}{2014}), \bibinfo{pages}{1}.
\newblock


\bibitem[\protect\citeauthoryear{Commission}{Commission}{2016}]%
        {ftc2016}
\bibfield{author}{\bibinfo{person}{Federal~Trade Commission}.}
  \bibinfo{year}{2016}\natexlab{}.
\newblock \bibinfo{title}{Big Data: A Tool for Inclusion or Exclusion?
  Understanding the Issues}.
\newblock
\newblock


\bibitem[\protect\citeauthoryear{Corden and Sainsbury}{Corden and
  Sainsbury}{2006}]%
        {corden2006verbatim}
\bibfield{author}{\bibinfo{person}{Anne Corden} {and} \bibinfo{person}{Roy
  Sainsbury}.} \bibinfo{year}{2006}\natexlab{}.
\newblock \bibinfo{booktitle}{\emph{Using verbatim quotations in reporting
  qualitative social research: researchers' views}}.
\newblock \bibinfo{publisher}{University of York}.
\newblock


\bibitem[\protect\citeauthoryear{Cottom}{Cottom}{2015}]%
        {cottom2015}
\bibfield{author}{\bibinfo{person}{Tressie~McMillan Cottom}.}
  \bibinfo{year}{2015}\natexlab{}.
\newblock \bibinfo{title}{Credit Scores, Life Chances, and Algorithms}.
\newblock
\newblock
\urldef\tempurl%
\url{https://tressiemc.com/uncategorized/credit-scores-life-chances-and-algorithms/}
\showURL{%
\tempurl}


\bibitem[\protect\citeauthoryear{De-Arteaga, Fogliato, and
  Chouldechova}{De-Arteaga et~al\mbox{.}}{2020}]%
        {dearteaga2020}
\bibfield{author}{\bibinfo{person}{Maria De-Arteaga}, \bibinfo{person}{Riccardo
  Fogliato}, {and} \bibinfo{person}{Alexandra Chouldechova}.}
  \bibinfo{year}{2020}\natexlab{}.
\newblock \showarticletitle{A Case for Humans-in-the-Loop: Decisions in the
  Presence of Erroneous Algorithmic Scores}. In
  \bibinfo{booktitle}{\emph{Proceedings of the 2020 CHI Conference on Human
  Factors in Computing Systems}}.
\newblock
\urldef\tempurl%
\url{https://doi.org/10.1145/3313831.3376638}
\showDOI{\tempurl}


\bibitem[\protect\citeauthoryear{Diakopoulos}{Diakopoulos}{2016}]%
        {diakopoulos2016}
\bibfield{author}{\bibinfo{person}{Nicholas Diakopoulos}.}
  \bibinfo{year}{2016}\natexlab{}.
\newblock \showarticletitle{Accountability in algorithmic decision making}.
\newblock \bibinfo{journal}{\emph{Commun. ACM}} \bibinfo{volume}{59},
  \bibinfo{number}{2} (\bibinfo{year}{2016}), \bibinfo{pages}{56--62}.
\newblock
\urldef\tempurl%
\url{https://doi.org/10.1145/2844110}
\showDOI{\tempurl}


\bibitem[\protect\citeauthoryear{DiSalvo, Lodato, Fries, Schechter, and
  Barnwell}{DiSalvo et~al\mbox{.}}{2011}]%
        {disalvo2011}
\bibfield{author}{\bibinfo{person}{Carl DiSalvo}, \bibinfo{person}{Thomas
  Lodato}, \bibinfo{person}{Laura Fries}, \bibinfo{person}{Beth Schechter},
  {and} \bibinfo{person}{Thomas Barnwell}.} \bibinfo{year}{2011}\natexlab{}.
\newblock \showarticletitle{The collective articulation of issues as design
  practice}.
\newblock \bibinfo{journal}{\emph{CoDesign}} \bibinfo{volume}{7},
  \bibinfo{number}{3-4} (\bibinfo{year}{2011}), \bibinfo{pages}{185--197}.
\newblock
\urldef\tempurl%
\url{https://doi.org/10.1080/15710882.2011.630475}
\showDOI{\tempurl}


\bibitem[\protect\citeauthoryear{DiSalvo, Nourbakhsh, Holstius, Akin, and
  Louw}{DiSalvo et~al\mbox{.}}{2008}]%
        {disalvo2008}
\bibfield{author}{\bibinfo{person}{Carl DiSalvo}, \bibinfo{person}{Illah
  Nourbakhsh}, \bibinfo{person}{David Holstius}, \bibinfo{person}{Ay{\c{c}}a
  Akin}, {and} \bibinfo{person}{Marti Louw}.} \bibinfo{year}{2008}\natexlab{}.
\newblock \showarticletitle{The Neighborhood Networks Project: A case study of
  critical engagement and creative expression through participatory design}. In
  \bibinfo{booktitle}{\emph{Proceedings of the Tenth Anniversary Conference on
  Participatory Design 2008 (PDC'08)}}. \bibinfo{pages}{41--50}.
\newblock


\bibitem[\protect\citeauthoryear{Doshi-Velez and Kim}{Doshi-Velez and
  Kim}{2017}]%
        {doshi2017}
\bibfield{author}{\bibinfo{person}{Finale Doshi-Velez} {and}
  \bibinfo{person}{Been Kim}.} \bibinfo{year}{2017}\natexlab{}.
\newblock \showarticletitle{Towards a rigorous science of interpretable machine
  learning}.
\newblock \bibinfo{journal}{\emph{arXiv preprint arXiv:1702.08608}}
  (\bibinfo{year}{2017}).
\newblock


\bibitem[\protect\citeauthoryear{Edelman}{Edelman}{2019}]%
        {edelman2019}
\bibfield{author}{\bibinfo{person}{Edelman}.} \bibinfo{year}{2019}\natexlab{}.
\newblock \bibinfo{title}{2019 Edelman AI Survey}.
\newblock
\newblock


\bibitem[\protect\citeauthoryear{Edwards and Veale}{Edwards and Veale}{2017}]%
        {edwards2017slave}
\bibfield{author}{\bibinfo{person}{Lilian Edwards} {and}
  \bibinfo{person}{Michael Veale}.} \bibinfo{year}{2017}\natexlab{}.
\newblock \showarticletitle{Slave to the algorithm? Why a right to an
  explanation is probably not the remedy you are looking for}.
\newblock \bibinfo{journal}{\emph{Duke Law and Technology Review}}
  \bibinfo{volume}{16} (\bibinfo{year}{2017}).
\newblock


\bibitem[\protect\citeauthoryear{Elish}{Elish}{2019}]%
        {elish2019}
\bibfield{author}{\bibinfo{person}{Madeleine~Clare Elish}.}
  \bibinfo{year}{2019}\natexlab{}.
\newblock \showarticletitle{Moral crumple zones: Cautionary tales in
  human-robot interaction}.
\newblock \bibinfo{journal}{\emph{Engaging Science, Technology, and Society}}
  \bibinfo{volume}{5} (\bibinfo{year}{2019}), \bibinfo{pages}{40--60}.
\newblock


\bibitem[\protect\citeauthoryear{Eslami, Karahalios, Sandvig, Vaccaro, Rickman,
  Hamilton, and Kirlik}{Eslami et~al\mbox{.}}{2016}]%
        {eslami2016}
\bibfield{author}{\bibinfo{person}{Motahhare Eslami}, \bibinfo{person}{Karrie
  Karahalios}, \bibinfo{person}{Christian Sandvig}, \bibinfo{person}{Kristen
  Vaccaro}, \bibinfo{person}{Aimee Rickman}, \bibinfo{person}{Kevin Hamilton},
  {and} \bibinfo{person}{Alex Kirlik}.} \bibinfo{year}{2016}\natexlab{}.
\newblock \showarticletitle{First I ``like" it, then I hide it: Folk Theories
  of Social Feeds}. In \bibinfo{booktitle}{\emph{Proceedings of the 2016 CHI
  Conference on Human Factors in Computing Systems}}.
  \bibinfo{pages}{2371--2382}.
\newblock
\urldef\tempurl%
\url{https://doi.org/10.1145/2858036.2858494}
\showDOI{\tempurl}


\bibitem[\protect\citeauthoryear{Eslami, Rickman, Vaccaro, Aleyasen, Vuong,
  Karahalios, Hamilton, and Sandvig}{Eslami et~al\mbox{.}}{2015}]%
        {eslami2015always}
\bibfield{author}{\bibinfo{person}{Motahhare Eslami}, \bibinfo{person}{Aimee
  Rickman}, \bibinfo{person}{Kristen Vaccaro}, \bibinfo{person}{Amirhossein
  Aleyasen}, \bibinfo{person}{Andy Vuong}, \bibinfo{person}{Karrie Karahalios},
  \bibinfo{person}{Kevin Hamilton}, {and} \bibinfo{person}{Christian Sandvig}.}
  \bibinfo{year}{2015}\natexlab{}.
\newblock \showarticletitle{"I always assumed that I wasn't really that close
  to [her]": Reasoning about Invisible Algorithms in News Feeds}. In
  \bibinfo{booktitle}{\emph{Proceedings of the 33rd Annual ACM Conference on
  Human Factors in Computing Systems}}. ACM, \bibinfo{pages}{153--162}.
\newblock
\urldef\tempurl%
\url{https://doi.org/10.1145/2702123.2702556}
\showDOI{\tempurl}


\bibitem[\protect\citeauthoryear{Eslami, Vaccaro, Lee, Elazari Bar~On, Gilbert,
  and Karahalios}{Eslami et~al\mbox{.}}{2019}]%
        {eslami2019}
\bibfield{author}{\bibinfo{person}{Motahhare Eslami}, \bibinfo{person}{Kristen
  Vaccaro}, \bibinfo{person}{Min~Kyung Lee}, \bibinfo{person}{Amit Elazari
  Bar~On}, \bibinfo{person}{Eric Gilbert}, {and} \bibinfo{person}{Karrie
  Karahalios}.} \bibinfo{year}{2019}\natexlab{}.
\newblock \showarticletitle{User attitudes towards algorithmic opacity and
  transparency in online reviewing platforms}. In
  \bibinfo{booktitle}{\emph{Proceedings of the 2019 CHI Conference on Human
  Factors in Computing Systems}}.
\newblock
\urldef\tempurl%
\url{https://doi.org/10.1145/3290605.3300724}
\showDOI{\tempurl}


\bibitem[\protect\citeauthoryear{Eubanks}{Eubanks}{2017}]%
        {eubanks2017}
\bibfield{author}{\bibinfo{person}{Virginia Eubanks}.}
  \bibinfo{year}{2017}\natexlab{}.
\newblock \bibinfo{booktitle}{\emph{Automating inequality: How high-tech tools
  profile, police, and punish the poor}}.
\newblock \bibinfo{publisher}{St. Martin's Press}.
\newblock


\bibitem[\protect\citeauthoryear{Fast and Horvitz}{Fast and Horvitz}{2017}]%
        {fast2017long}
\bibfield{author}{\bibinfo{person}{Ethan Fast} {and} \bibinfo{person}{Eric
  Horvitz}.} \bibinfo{year}{2017}\natexlab{}.
\newblock \showarticletitle{Long-term trends in the public perception of
  artificial intelligence}. In \bibinfo{booktitle}{\emph{Thirty-First AAAI
  Conference on Artificial Intelligence}}.
\newblock


\bibitem[\protect\citeauthoryear{Fjeld, Achten, Hilligoss, Nagy, and
  Srikumar}{Fjeld et~al\mbox{.}}{2020}]%
        {fjeld2020}
\bibfield{author}{\bibinfo{person}{Jessica Fjeld}, \bibinfo{person}{Nele
  Achten}, \bibinfo{person}{Hannah Hilligoss}, \bibinfo{person}{Adam Nagy},
  {and} \bibinfo{person}{Madhulika Srikumar}.} \bibinfo{year}{2020}\natexlab{}.
\newblock \showarticletitle{Principled artificial intelligence: Mapping
  consensus in ethical and rights-based approaches to principles for AI}.
\newblock \bibinfo{journal}{\emph{Berkman Klein Center Research Publication}}
  \bibinfo{number}{2020-1} (\bibinfo{year}{2020}).
\newblock


\bibitem[\protect\citeauthoryear{Fourcade and Healy}{Fourcade and
  Healy}{2013}]%
        {fourcade2013}
\bibfield{author}{\bibinfo{person}{Marion Fourcade} {and}
  \bibinfo{person}{Kieran Healy}.} \bibinfo{year}{2013}\natexlab{}.
\newblock \showarticletitle{Classification situations: Life-chances in the
  neoliberal era}.
\newblock \bibinfo{journal}{\emph{Accounting, Organizations and Society}}
  \bibinfo{volume}{38}, \bibinfo{number}{8} (\bibinfo{year}{2013}),
  \bibinfo{pages}{559--572}.
\newblock


\bibitem[\protect\citeauthoryear{Friedman and Nissenbaum}{Friedman and
  Nissenbaum}{1996}]%
        {friedman1996}
\bibfield{author}{\bibinfo{person}{Batya Friedman} {and} \bibinfo{person}{Helen
  Nissenbaum}.} \bibinfo{year}{1996}\natexlab{}.
\newblock \showarticletitle{Bias in computer systems}.
\newblock \bibinfo{journal}{\emph{ACM Transactions on Information Systems
  (TOIS)}} \bibinfo{volume}{14}, \bibinfo{number}{3} (\bibinfo{year}{1996}),
  \bibinfo{pages}{330--347}.
\newblock
\urldef\tempurl%
\url{https://doi.org/10.1145/230538.230561}
\showDOI{\tempurl}


\bibitem[\protect\citeauthoryear{Garvey and Maskal}{Garvey and Maskal}{2019}]%
        {garvey2019sentiment}
\bibfield{author}{\bibinfo{person}{Colin Garvey} {and}
  \bibinfo{person}{Chandler Maskal}.} \bibinfo{year}{2019}\natexlab{}.
\newblock \showarticletitle{Sentiment Analysis of the News Media on Artificial
  Intelligence Does Not Support Claims of Negative Bias Against Artificial
  Intelligence}.
\newblock \bibinfo{journal}{\emph{OMICS: a Journal of Integrative Biology}}
  (\bibinfo{year}{2019}).
\newblock


\bibitem[\protect\citeauthoryear{Gaver, Beaver, and Benford}{Gaver
  et~al\mbox{.}}{2003}]%
        {gaver2003}
\bibfield{author}{\bibinfo{person}{William~W. Gaver}, \bibinfo{person}{Jacob
  Beaver}, {and} \bibinfo{person}{Steve Benford}.}
  \bibinfo{year}{2003}\natexlab{}.
\newblock \showarticletitle{Ambiguity as a resource for design}. In
  \bibinfo{booktitle}{\emph{Proceedings of the SIGCHI Conference on Human
  Factors in Computing Systems}}. \bibinfo{pages}{233--240}.
\newblock
\urldef\tempurl%
\url{https://doi.org/10.1145/642611.642653}
\showDOI{\tempurl}


\bibitem[\protect\citeauthoryear{Gillespie}{Gillespie}{2014}]%
        {gillespie2014relevance}
\bibfield{author}{\bibinfo{person}{Tarleton Gillespie}.}
  \bibinfo{year}{2014}\natexlab{}.
\newblock \bibinfo{booktitle}{\emph{The relevance of algorithms}}.
\newblock \bibinfo{publisher}{MIT Press}.
\newblock


\bibitem[\protect\citeauthoryear{Gilpin, Bau, Yuan, Bajwa, Specter, and
  Kagal}{Gilpin et~al\mbox{.}}{2018}]%
        {gilpin2018}
\bibfield{author}{\bibinfo{person}{Leilani~H. Gilpin}, \bibinfo{person}{David
  Bau}, \bibinfo{person}{Ben~Z. Yuan}, \bibinfo{person}{Ayesha Bajwa},
  \bibinfo{person}{Michael Specter}, {and} \bibinfo{person}{Lalana Kagal}.}
  \bibinfo{year}{2018}\natexlab{}.
\newblock \showarticletitle{Explaining explanations: An overview of
  interpretability of machine learning}. In \bibinfo{booktitle}{\emph{2018 IEEE
  5th International Conference on Data Science and Advanced Analytics (DSAA)}}.
  IEEE, \bibinfo{pages}{80--89}.
\newblock


\bibitem[\protect\citeauthoryear{Goodman and Flaxman}{Goodman and
  Flaxman}{2017}]%
        {goodman2017european}
\bibfield{author}{\bibinfo{person}{Bryce Goodman} {and} \bibinfo{person}{Seth
  Flaxman}.} \bibinfo{year}{2017}\natexlab{}.
\newblock \showarticletitle{European Union regulations on algorithmic
  decision-making and a “right to explanation”}.
\newblock \bibinfo{journal}{\emph{AI magazine}} \bibinfo{volume}{38},
  \bibinfo{number}{3} (\bibinfo{year}{2017}), \bibinfo{pages}{50--57}.
\newblock


\bibitem[\protect\citeauthoryear{Green and Chen}{Green and Chen}{2019}]%
        {green2019}
\bibfield{author}{\bibinfo{person}{Ben Green} {and} \bibinfo{person}{Yiling
  Chen}.} \bibinfo{year}{2019}\natexlab{}.
\newblock \showarticletitle{The Principles and Limits of Algorithm-in-the-Loop
  Decision Making}.
\newblock  \bibinfo{volume}{3}, \bibinfo{number}{CSCW} (\bibinfo{year}{2019}).
\newblock
\urldef\tempurl%
\url{https://doi.org/10.1145/3359152}
\showDOI{\tempurl}


\bibitem[\protect\citeauthoryear{Gupta, Cotter, Pfeifer, Voevodski, Canini,
  Mangylov, Moczydlowski, and Van~Esbroeck}{Gupta et~al\mbox{.}}{2016}]%
        {gupta2016}
\bibfield{author}{\bibinfo{person}{Maya Gupta}, \bibinfo{person}{Andrew
  Cotter}, \bibinfo{person}{Jan Pfeifer}, \bibinfo{person}{Konstantin
  Voevodski}, \bibinfo{person}{Kevin Canini}, \bibinfo{person}{Alexander
  Mangylov}, \bibinfo{person}{Wojciech Moczydlowski}, {and}
  \bibinfo{person}{Alexander Van~Esbroeck}.} \bibinfo{year}{2016}\natexlab{}.
\newblock \showarticletitle{Monotonic calibrated interpolated look-up tables}.
\newblock \bibinfo{journal}{\emph{The Journal of Machine Learning Research}}
  \bibinfo{volume}{17}, \bibinfo{number}{109} (\bibinfo{year}{2016}),
  \bibinfo{pages}{1--47}.
\newblock


\bibitem[\protect\citeauthoryear{Herman}{Herman}{2017}]%
        {herman2017}
\bibfield{author}{\bibinfo{person}{Bernease Herman}.}
  \bibinfo{year}{2017}\natexlab{}.
\newblock \showarticletitle{The promise and peril of human evaluation for model
  interpretability}. In \bibinfo{booktitle}{\emph{Proceedings of the 31st
  Conference on Neural Information Processing Systems (NIPS 2017)}}.
\newblock


\bibitem[\protect\citeauthoryear{Hirsch, Merced, Narayanan, Imel, and
  Atkins}{Hirsch et~al\mbox{.}}{2017}]%
        {hirsch2017}
\bibfield{author}{\bibinfo{person}{Tad Hirsch}, \bibinfo{person}{Kritzia
  Merced}, \bibinfo{person}{Shrikanth Narayanan}, \bibinfo{person}{Zac~E.
  Imel}, {and} \bibinfo{person}{David~C. Atkins}.}
  \bibinfo{year}{2017}\natexlab{}.
\newblock \showarticletitle{Designing contestability: Interaction design,
  machine learning, and mental health}. In
  \bibinfo{booktitle}{\emph{Proceedings of the 2017 Conference on Designing
  Interactive Systems}}. \bibinfo{pages}{95--99}.
\newblock
\urldef\tempurl%
\url{https://doi.org/10.1145/3064663.3064703}
\showDOI{\tempurl}


\bibitem[\protect\citeauthoryear{Immonen, Sintonen, and Koivuniemi}{Immonen
  et~al\mbox{.}}{2018}]%
        {immonen2018}
\bibfield{author}{\bibinfo{person}{Mika Immonen}, \bibinfo{person}{Sanna
  Sintonen}, {and} \bibinfo{person}{Jouni Koivuniemi}.}
  \bibinfo{year}{2018}\natexlab{}.
\newblock \showarticletitle{The value of human interaction in service
  channels}.
\newblock \bibinfo{journal}{\emph{Computers in Human Behavior}}
  \bibinfo{volume}{78} (\bibinfo{year}{2018}), \bibinfo{pages}{316--325}.
\newblock


\bibitem[\protect\citeauthoryear{Introna and Nissenbaum}{Introna and
  Nissenbaum}{2000}]%
        {introna2000}
\bibfield{author}{\bibinfo{person}{Lucas~D. Introna} {and}
  \bibinfo{person}{Helen Nissenbaum}.} \bibinfo{year}{2000}\natexlab{}.
\newblock \showarticletitle{Shaping the Web: Why the politics of search engines
  matters}.
\newblock \bibinfo{journal}{\emph{The Information Society}}
  \bibinfo{volume}{16}, \bibinfo{number}{3} (\bibinfo{year}{2000}),
  \bibinfo{pages}{169--185}.
\newblock
\urldef\tempurl%
\url{https://doi.org/10.1080/01972240050133634}
\showDOI{\tempurl}


\bibitem[\protect\citeauthoryear{Ipsos}{Ipsos}{2019}]%
        {ipsos2019}
\bibfield{author}{\bibinfo{person}{Ipsos}.} \bibinfo{year}{2019}\natexlab{}.
\newblock \bibinfo{title}{Widespread concern about artificial intelligence}.
\newblock
\newblock
\urldef\tempurl%
\url{www.ipsos.com/en/widespread-concern-about-artificial-intelligence/}
\showURL{%
\tempurl}


\bibitem[\protect\citeauthoryear{Jungk and M{\"u}llert}{Jungk and
  M{\"u}llert}{1987}]%
        {jungk1987}
\bibfield{author}{\bibinfo{person}{Robert Jungk} {and} \bibinfo{person}{Norbert
  M{\"u}llert}.} \bibinfo{year}{1987}\natexlab{}.
\newblock \bibinfo{booktitle}{\emph{Future Workshops: How to create desirable
  futures}}.
\newblock \bibinfo{publisher}{Institute for Social Inventions}.
\newblock


\bibitem[\protect\citeauthoryear{Kaminski}{Kaminski}{2019a}]%
        {kaminski2019binary}
\bibfield{author}{\bibinfo{person}{Margot~E. Kaminski}.}
  \bibinfo{year}{2019}\natexlab{a}.
\newblock \showarticletitle{Binary Governance: Lessons from the GDPR's Approach
  to Algorithmic Accountability}.
\newblock \bibinfo{journal}{\emph{Southern California Law Review}}
  \bibinfo{volume}{92}, \bibinfo{number}{6} (\bibinfo{year}{2019}).
\newblock


\bibitem[\protect\citeauthoryear{Kaminski}{Kaminski}{2019b}]%
        {kaminski2019right}
\bibfield{author}{\bibinfo{person}{Margot~E. Kaminski}.}
  \bibinfo{year}{2019}\natexlab{b}.
\newblock \showarticletitle{The right to explanation, explained}.
\newblock \bibinfo{journal}{\emph{Berkeley Technology Law Journal}}
  \bibinfo{volume}{34}, \bibinfo{number}{1} (\bibinfo{year}{2019}).
\newblock


\bibitem[\protect\citeauthoryear{Karimi, Barthe, Balle, and Valera}{Karimi
  et~al\mbox{.}}{2020}]%
        {karimi2020}
\bibfield{author}{\bibinfo{person}{Amir-Hossein Karimi},
  \bibinfo{person}{Gilles Barthe}, \bibinfo{person}{Borja Balle}, {and}
  \bibinfo{person}{Isabel Valera}.} \bibinfo{year}{2020}\natexlab{}.
\newblock \showarticletitle{Model-Agnostic Counterfactual Explanations for
  Consequential Decisions}. In \bibinfo{booktitle}{\emph{Proceedings of the
  23rd International Conference on Artificial Intelligence and Statistics
  (AISTATS)}}.
\newblock


\bibitem[\protect\citeauthoryear{Kelley, Yang, Heldreth, Moessner, Sedley,
  Kramm, Newman, and Woodruff}{Kelley et~al\mbox{.}}{2020}]%
        {kelley2020}
\bibfield{author}{\bibinfo{person}{Patrick~Gage Kelley},
  \bibinfo{person}{Yongwei Yang}, \bibinfo{person}{Courtney Heldreth},
  \bibinfo{person}{Christopher Moessner}, \bibinfo{person}{Aaron Sedley},
  \bibinfo{person}{Andreas Kramm}, \bibinfo{person}{David Newman}, {and}
  \bibinfo{person}{Allison Woodruff}.} \bibinfo{year}{2020}\natexlab{}.
\newblock \showarticletitle{``Happy and Assured that life will be easy 10years
  from now.'': Perceptions of Artificial Intelligence in 8 Countries}.
\newblock \bibinfo{journal}{\emph{arXiv preprint arXiv:2001.00081}}
  (\bibinfo{year}{2020}).
\newblock


\bibitem[\protect\citeauthoryear{Kluttz, Kohli, and Mulligan}{Kluttz
  et~al\mbox{.}}{2020}]%
        {kluttz2019}
\bibfield{author}{\bibinfo{person}{Daniel Kluttz}, \bibinfo{person}{Nitin
  Kohli}, {and} \bibinfo{person}{Deirdre~K. Mulligan}.}
  \bibinfo{year}{2020}\natexlab{}.
\newblock \showarticletitle{Shaping Our Tools: Contestability as a Means to
  Promote Responsible Algorithmic Decision Making in the Professions}. In
  \bibinfo{booktitle}{\emph{After the Digital Tornado: Networks, Algorithms,
  Humanity}}, \bibfield{editor}{\bibinfo{person}{Kevin Werbach}} (Ed.).
  \bibinfo{publisher}{Cambridge University Press}.
\newblock


\bibitem[\protect\citeauthoryear{Kroll, Barocas, Felten, Reidenberg, Robinson,
  and Yu}{Kroll et~al\mbox{.}}{2016}]%
        {kroll2016}
\bibfield{author}{\bibinfo{person}{Joshua~A. Kroll}, \bibinfo{person}{Solon
  Barocas}, \bibinfo{person}{Edward~W. Felten}, \bibinfo{person}{Joel~R.
  Reidenberg}, \bibinfo{person}{David~G. Robinson}, {and}
  \bibinfo{person}{Harlan Yu}.} \bibinfo{year}{2016}\natexlab{}.
\newblock \showarticletitle{Accountable algorithms}.
\newblock \bibinfo{journal}{\emph{University of Pennsylvania Law Review}}
  \bibinfo{volume}{165} (\bibinfo{year}{2016}).
\newblock


\bibitem[\protect\citeauthoryear{Lipton}{Lipton}{2016}]%
        {lipton2016}
\bibfield{author}{\bibinfo{person}{Zachary~C. Lipton}.}
  \bibinfo{year}{2016}\natexlab{}.
\newblock \showarticletitle{The mythos of model interpretability}. In
  \bibinfo{booktitle}{\emph{2016 ICML Workshop on Human Interpretability in
  Machine Learning (WHI 2016)}}.
\newblock


\bibitem[\protect\citeauthoryear{London}{London}{2019}]%
        {london2019}
\bibfield{author}{\bibinfo{person}{Alex~John London}.}
  \bibinfo{year}{2019}\natexlab{}.
\newblock \showarticletitle{Artificial intelligence and black-box medical
  decisions: accuracy versus explainability}.
\newblock \bibinfo{journal}{\emph{Hastings Center Report}}
  \bibinfo{volume}{49}, \bibinfo{number}{1} (\bibinfo{year}{2019}),
  \bibinfo{pages}{15--21}.
\newblock


\bibitem[\protect\citeauthoryear{Lustig, Pine, Nardi, Irani, Lee, Nafus, and
  Sandvig}{Lustig et~al\mbox{.}}{2016}]%
        {lustig2016}
\bibfield{author}{\bibinfo{person}{Caitlin Lustig}, \bibinfo{person}{Katie
  Pine}, \bibinfo{person}{Bonnie Nardi}, \bibinfo{person}{Lilly Irani},
  \bibinfo{person}{Min~Kyung Lee}, \bibinfo{person}{Dawn Nafus}, {and}
  \bibinfo{person}{Christian Sandvig}.} \bibinfo{year}{2016}\natexlab{}.
\newblock \showarticletitle{Algorithmic authority: the ethics, politics, and
  economics of algorithms that interpret, decide, and manage}. In
  \bibinfo{booktitle}{\emph{Proceedings of the 2016 CHI Conference Extended
  Abstracts on Human Factors in Computing Systems}}.
  \bibinfo{pages}{1057--1062}.
\newblock
\urldef\tempurl%
\url{https://doi.org/10.1145/2851581.2886426}
\showDOI{\tempurl}


\bibitem[\protect\citeauthoryear{Miller}{Miller}{2019}]%
        {miller2019}
\bibfield{author}{\bibinfo{person}{Tim Miller}.}
  \bibinfo{year}{2019}\natexlab{}.
\newblock \showarticletitle{Explanation in artificial intelligence: Insights
  from the social sciences}.
\newblock \bibinfo{journal}{\emph{Artificial Intelligence}}
  \bibinfo{volume}{267} (\bibinfo{year}{2019}), \bibinfo{pages}{1--38}.
\newblock


\bibitem[\protect\citeauthoryear{MORI}{MORI}{2017}]%
        {ipsosmori2017}
\bibfield{author}{\bibinfo{person}{Ipsos MORI}.}
  \bibinfo{year}{2017}\natexlab{}.
\newblock \bibinfo{title}{Public views of Machine Learning: Findings from
  public research and engagement conducted on behalf of the Royal Society}.
\newblock
\newblock
\urldef\tempurl%
\url{https://royalsociety.org/-/media/policy/projects/machine-learning/publications/public-views-of-machine-learning-ipsos-mori.pdf}
\showURL{%
\tempurl}


\bibitem[\protect\citeauthoryear{Mozilla}{Mozilla}{2019}]%
        {mozilla2019}
\bibfield{author}{\bibinfo{person}{Mozilla}.} \bibinfo{year}{2019}\natexlab{}.
\newblock \bibinfo{title}{We Asked People Around the World How They Feel About
  Artificial Intelligence. Here's What We Learned.}
\newblock
\newblock
\urldef\tempurl%
\url{https://foundation.mozilla.org/en/blog/we-asked-people-around-the-world-how-they-feel-about-artificial-intelligence-heres-what-we-learned/}
\showURL{%
\tempurl}


\bibitem[\protect\citeauthoryear{Munoz, Smith, and Patil}{Munoz
  et~al\mbox{.}}{2016}]%
        {executive2016}
\bibfield{author}{\bibinfo{person}{Cecilia Munoz}, \bibinfo{person}{Megan
  Smith}, {and} \bibinfo{person}{D.J. Patil}.} \bibinfo{year}{2016}\natexlab{}.
\newblock \bibinfo{booktitle}{\emph{Big data: A report on algorithmic systems,
  opportunity, and civil rights}}.
\newblock \bibinfo{publisher}{Executive Office of the President}.
\newblock


\bibitem[\protect\citeauthoryear{Northstar}{Northstar}{2017}]%
        {arm2017}
\bibfield{author}{\bibinfo{person}{ARM~| Northstar}.}
  \bibinfo{year}{2017}\natexlab{}.
\newblock \bibinfo{title}{AI Today, AI Tomorrow. Awareness and Anticipation of
  AI: A Global Perspective}.
\newblock
\newblock
\urldef\tempurl%
\url{www.arm.com/solutions/artificial-intelligence/survey/}
\showURL{%
\tempurl}


\bibitem[\protect\citeauthoryear{Nunes and Jannach}{Nunes and Jannach}{2017}]%
        {nunes2017}
\bibfield{author}{\bibinfo{person}{Ingrid Nunes} {and} \bibinfo{person}{Dietmar
  Jannach}.} \bibinfo{year}{2017}\natexlab{}.
\newblock \showarticletitle{A systematic review and taxonomy of explanations in
  decision support and recommender systems}.
\newblock \bibinfo{journal}{\emph{User Modeling and User-Adapted Interaction}}
  \bibinfo{volume}{27}, \bibinfo{number}{3-5} (\bibinfo{year}{2017}),
  \bibinfo{pages}{393--444}.
\newblock


\bibitem[\protect\citeauthoryear{Office}{Office}{2019}]%
        {ico2019}
\bibfield{author}{\bibinfo{person}{Information~Commissioner’s Office}.}
  \bibinfo{year}{2019}\natexlab{}.
\newblock \bibinfo{title}{Project ExplAIn Interim Report}.
\newblock
\newblock
\urldef\tempurl%
\url{https://ico.org.uk/about-the-ico/research-and-reports/project-explain-interim-report/}
\showURL{%
\tempurl}


\bibitem[\protect\citeauthoryear{on~Artificial~Intelligence}{on~Artificial~Intelligence}{2019}]%
        {highlevel2019}
\bibfield{author}{\bibinfo{person}{High-Level Expert~Group on
  Artificial~Intelligence}.} \bibinfo{year}{2019}\natexlab{}.
\newblock \bibinfo{title}{Ethics Guidelines for Trustworthy AI}.
\newblock
\newblock


\bibitem[\protect\citeauthoryear{O'Neil}{O'Neil}{2016}]%
        {oneil2016}
\bibfield{author}{\bibinfo{person}{Cathy O'Neil}.}
  \bibinfo{year}{2016}\natexlab{}.
\newblock \bibinfo{booktitle}{\emph{Weapons of math destruction: How big data
  increases inequality and threatens democracy}}.
\newblock \bibinfo{publisher}{Broadway Books}.
\newblock


\bibitem[\protect\citeauthoryear{Ostrom, Bitner, and Meuter}{Ostrom
  et~al\mbox{.}}{2002}]%
        {ostrom2002}
\bibfield{author}{\bibinfo{person}{Amy~L. Ostrom}, \bibinfo{person}{Mary~J.
  Bitner}, {and} \bibinfo{person}{Matthew~L. Meuter}.}
  \bibinfo{year}{2002}\natexlab{}.
\newblock \showarticletitle{Self-service technologies}.
\newblock \bibinfo{journal}{\emph{e-Service: New Directions in Theory and
  Practice, ME Sharpe, NY}} (\bibinfo{year}{2002}), \bibinfo{pages}{45--64}.
\newblock


\bibitem[\protect\citeauthoryear{Oswald}{Oswald}{2019}]%
        {oswald2019}
\bibfield{author}{\bibinfo{person}{Malcolm Oswald}.}
  \bibinfo{year}{2019}\natexlab{}.
\newblock \bibinfo{title}{Artificial Intelligence (AI) \& Explainability
  Citizens’ Juries Report}.
\newblock
\newblock
\urldef\tempurl%
\url{https://assets.mhs.manchester.ac.uk/gmpstrc/C4-AI-citizens-juries-report.pdf}
\showURL{%
\tempurl}


\bibitem[\protect\citeauthoryear{Parasuraman, Sheridan, and
  Wickens}{Parasuraman et~al\mbox{.}}{2000}]%
        {parasuraman2000}
\bibfield{author}{\bibinfo{person}{Raja Parasuraman},
  \bibinfo{person}{Thomas~B. Sheridan}, {and} \bibinfo{person}{Christopher~D.
  Wickens}.} \bibinfo{year}{2000}\natexlab{}.
\newblock \showarticletitle{A model for types and levels of human interaction
  with automation}.
\newblock \bibinfo{journal}{\emph{IEEE Transactions on Systems, Man, and
  Cybernetics - Part A: Systems and Humans}} \bibinfo{volume}{30},
  \bibinfo{number}{3} (\bibinfo{year}{2000}), \bibinfo{pages}{286--297}.
\newblock


\bibitem[\protect\citeauthoryear{Pasquale}{Pasquale}{2015}]%
        {pasquale2015}
\bibfield{author}{\bibinfo{person}{Frank Pasquale}.}
  \bibinfo{year}{2015}\natexlab{}.
\newblock \bibinfo{booktitle}{\emph{The Black Box Society: The Secret
  Algorithms that Control Money and Information}}.
\newblock \bibinfo{publisher}{Harvard University Press}.
\newblock


\bibitem[\protect\citeauthoryear{Poursabzi-Sangdeh, Goldstein, Hofman, Vaughan,
  and Wallach}{Poursabzi-Sangdeh et~al\mbox{.}}{2018}]%
        {poursabzi2018}
\bibfield{author}{\bibinfo{person}{Forough Poursabzi-Sangdeh},
  \bibinfo{person}{Daniel~G. Goldstein}, \bibinfo{person}{Jake~M. Hofman},
  \bibinfo{person}{Jennifer~Wortman Vaughan}, {and} \bibinfo{person}{Hanna
  Wallach}.} \bibinfo{year}{2018}\natexlab{}.
\newblock \showarticletitle{Manipulating and measuring model interpretability}.
\newblock \bibinfo{journal}{\emph{arXiv preprint arXiv:1802.07810}}
  (\bibinfo{year}{2018}).
\newblock


\bibitem[\protect\citeauthoryear{Rader, Cotter, and Cho}{Rader
  et~al\mbox{.}}{2018}]%
        {rader2018}
\bibfield{author}{\bibinfo{person}{Emilee Rader}, \bibinfo{person}{Kelley
  Cotter}, {and} \bibinfo{person}{Janghee Cho}.}
  \bibinfo{year}{2018}\natexlab{}.
\newblock \showarticletitle{Explanations as Mechanisms for Supporting
  Algorithmic Transparency}. In \bibinfo{booktitle}{\emph{Proceedings of the
  2018 CHI Conference on Human Factors in Computing Systems}}.
\newblock
\urldef\tempurl%
\url{https://doi.org/10.1145/3173574.3173677}
\showDOI{\tempurl}


\bibitem[\protect\citeauthoryear{Rader and Gray}{Rader and Gray}{2015}]%
        {rader2015understanding}
\bibfield{author}{\bibinfo{person}{Emilee Rader} {and} \bibinfo{person}{Rebecca
  Gray}.} \bibinfo{year}{2015}\natexlab{}.
\newblock \showarticletitle{Understanding User Beliefs about Algorithmic
  Curation in the Facebook News Feed}. In \bibinfo{booktitle}{\emph{Proceedings
  of the 33rd Annual ACM Conference on Human Factors in Computing Systems}}.
  \bibinfo{pages}{173--182}.
\newblock
\urldef\tempurl%
\url{https://doi.org/10.1145/2702123.2702174}
\showDOI{\tempurl}


\bibitem[\protect\citeauthoryear{Ribeiro, Singh, and Guestrin}{Ribeiro
  et~al\mbox{.}}{2016}]%
        {ribeiro2016}
\bibfield{author}{\bibinfo{person}{Marco~Tulio Ribeiro},
  \bibinfo{person}{Sameer Singh}, {and} \bibinfo{person}{Carlos Guestrin}.}
  \bibinfo{year}{2016}\natexlab{}.
\newblock \showarticletitle{`Why should I trust you?': Explaining the
  predictions of any classifier}. In \bibinfo{booktitle}{\emph{Proceedings of
  the 22nd ACM SIGKDD International Conference on Knowledge Discovery and Data
  Mining}}. \bibinfo{pages}{1135--1144}.
\newblock


\bibitem[\protect\citeauthoryear{Rosner, Kawas, Li, Tilly, and Sung}{Rosner
  et~al\mbox{.}}{2016}]%
        {rosner2016}
\bibfield{author}{\bibinfo{person}{Daniela~K. Rosner}, \bibinfo{person}{Saba
  Kawas}, \bibinfo{person}{Wenqi Li}, \bibinfo{person}{Nicole Tilly}, {and}
  \bibinfo{person}{Yi-Chen Sung}.} \bibinfo{year}{2016}\natexlab{}.
\newblock \showarticletitle{Out of time, out of place: Reflections on design
  workshops as a research method}. In \bibinfo{booktitle}{\emph{Proceedings of
  the 19th ACM Conference on Computer-Supported Cooperative Work \& Social
  Computing (CSCW '16)}}. \bibinfo{pages}{1131--1141}.
\newblock
\urldef\tempurl%
\url{https://doi.org/10.1145/2818048.2820021}
\showDOI{\tempurl}


\bibitem[\protect\citeauthoryear{Rudin}{Rudin}{2019}]%
        {rudin2019}
\bibfield{author}{\bibinfo{person}{Cynthia Rudin}.}
  \bibinfo{year}{2019}\natexlab{}.
\newblock \showarticletitle{Stop explaining black box machine learning models
  for high stakes decisions and use interpretable models instead}.
\newblock \bibinfo{journal}{\emph{Nature Machine Intelligence}}
  \bibinfo{volume}{1}, \bibinfo{number}{5} (\bibinfo{year}{2019}),
  \bibinfo{pages}{206--215}.
\newblock


\bibitem[\protect\citeauthoryear{Russell}{Russell}{2019}]%
        {russell2020}
\bibfield{author}{\bibinfo{person}{Chris Russell}.}
  \bibinfo{year}{2019}\natexlab{}.
\newblock \showarticletitle{Efficient Search for Diverse Coherent
  Explanations}. In \bibinfo{booktitle}{\emph{Proceedings of the Conference on
  Fairness, Accountability, and Transparency}} \emph{(\bibinfo{series}{FAT*
  '19})}. \bibinfo{pages}{20–28}.
\newblock
\urldef\tempurl%
\url{https://doi.org/10.1145/3287560.3287569}
\showDOI{\tempurl}


\bibitem[\protect\citeauthoryear{Sandvig, Hamilton, Karahalios, and
  Langbort}{Sandvig et~al\mbox{.}}{2015}]%
        {sandvig2015}
\bibfield{author}{\bibinfo{person}{Christian Sandvig}, \bibinfo{person}{Kevin
  Hamilton}, \bibinfo{person}{Karrie Karahalios}, {and} \bibinfo{person}{Cedric
  Langbort}.} \bibinfo{year}{2015}\natexlab{}.
\newblock \showarticletitle{Can an Algorithm be Unethical?}. In
  \bibinfo{booktitle}{\emph{65th Annual Meeting of the International
  Communication Association}}.
\newblock


\bibitem[\protect\citeauthoryear{Selbst and Barocas}{Selbst and
  Barocas}{2018}]%
        {selbst2018intuitive}
\bibfield{author}{\bibinfo{person}{Andrew~D. Selbst} {and}
  \bibinfo{person}{Solon Barocas}.} \bibinfo{year}{2018}\natexlab{}.
\newblock \showarticletitle{The intuitive appeal of explainable machines}.
\newblock \bibinfo{journal}{\emph{Fordham Law Review}}  \bibinfo{volume}{87}
  (\bibinfo{year}{2018}), \bibinfo{pages}{1085--1139}.
\newblock


\bibitem[\protect\citeauthoryear{Sendak, Elish, Gao, Futoma, Ratliff, Nichols,
  Bedoya, Balu, and O'Brien}{Sendak et~al\mbox{.}}{2020}]%
        {sendak2020}
\bibfield{author}{\bibinfo{person}{Mark Sendak},
  \bibinfo{person}{Madeleine~Clare Elish}, \bibinfo{person}{Michael Gao},
  \bibinfo{person}{Joseph Futoma}, \bibinfo{person}{William Ratliff},
  \bibinfo{person}{Marshall Nichols}, \bibinfo{person}{Armando Bedoya},
  \bibinfo{person}{Suresh Balu}, {and} \bibinfo{person}{Cara O'Brien}.}
  \bibinfo{year}{2020}\natexlab{}.
\newblock \showarticletitle{`The human body is a black box': supporting
  clinical decision-making with deep learning}. In
  \bibinfo{booktitle}{\emph{Proceedings of the 2020 Conference on Fairness,
  Accountability, and Transparency}}. \bibinfo{pages}{99--109}.
\newblock
\urldef\tempurl%
\url{https://doi.org/10.1145/3351095.3372827}
\showDOI{\tempurl}


\bibitem[\protect\citeauthoryear{Singh}{Singh}{2019}]%
        {singh2019}
\bibfield{author}{\bibinfo{person}{Asheem Singh}.}
  \bibinfo{year}{2019}\natexlab{}.
\newblock \showarticletitle{Democratising decisions about technology: A
  toolkit}.
\newblock  (\bibinfo{year}{2019}).
\newblock
\urldef\tempurl%
\url{https://www.thersa.org/discover/publications-and-articles/reports/democratising-decisions-technology-toolkit}
\showURL{%
\tempurl}


\bibitem[\protect\citeauthoryear{Thomas}{Thomas}{2006}]%
        {thomas2006inductive}
\bibfield{author}{\bibinfo{person}{David~R. Thomas}.}
  \bibinfo{year}{2006}\natexlab{}.
\newblock \showarticletitle{A general inductive approach for analyzing
  qualitative evaluation data}.
\newblock \bibinfo{journal}{\emph{American Journal of Evaluation}}
  \bibinfo{volume}{27}, \bibinfo{number}{2} (\bibinfo{year}{2006}),
  \bibinfo{pages}{237--246}.
\newblock


\bibitem[\protect\citeauthoryear{Tintarev and Masthoff}{Tintarev and
  Masthoff}{2007}]%
        {tintarev2007}
\bibfield{author}{\bibinfo{person}{Nava Tintarev} {and} \bibinfo{person}{Judith
  Masthoff}.} \bibinfo{year}{2007}\natexlab{}.
\newblock \showarticletitle{A survey of explanations in recommender systems}.
  In \bibinfo{booktitle}{\emph{Proceedings of the 2007 IEEE 23rd International
  Conference on Data Engineering Workshop}}. \bibinfo{pages}{801--810}.
\newblock


\bibitem[\protect\citeauthoryear{University and Gallup}{University and
  Gallup}{2018}]%
        {northeastern2018}
\bibfield{author}{\bibinfo{person}{Northeastern University} {and}
  \bibinfo{person}{Gallup}.} \bibinfo{year}{2018}\natexlab{}.
\newblock \bibinfo{title}{Optimism and Anxiety: Views on the Impact of
  Artificial Intelligence and Higher Education's Response}.
\newblock
\newblock


\bibitem[\protect\citeauthoryear{Ur, Leon, Cranor, Shay, and Wang}{Ur
  et~al\mbox{.}}{2012}]%
        {ur2012smart}
\bibfield{author}{\bibinfo{person}{Blase Ur}, \bibinfo{person}{Pedro~Giovanni
  Leon}, \bibinfo{person}{Lorrie~Faith Cranor}, \bibinfo{person}{Richard Shay},
  {and} \bibinfo{person}{Yang Wang}.} \bibinfo{year}{2012}\natexlab{}.
\newblock \showarticletitle{Smart, useful, scary, creepy: perceptions of online
  behavioral advertising}. In \bibinfo{booktitle}{\emph{Proceedings of the
  Eighth Symposium on Usable Privacy and Security (SOUPS) 2012}}.
\newblock
\urldef\tempurl%
\url{https://doi.org/10.1145/2335356.2335362}
\showDOI{\tempurl}


\bibitem[\protect\citeauthoryear{Ustun, Spangher, and Liu}{Ustun
  et~al\mbox{.}}{2019}]%
        {ustun2019}
\bibfield{author}{\bibinfo{person}{Berk Ustun}, \bibinfo{person}{Alexander
  Spangher}, {and} \bibinfo{person}{Yang Liu}.}
  \bibinfo{year}{2019}\natexlab{}.
\newblock \showarticletitle{Actionable Recourse in Linear Classification}. In
  \bibinfo{booktitle}{\emph{Proceedings of the Conference on Fairness,
  Accountability, and Transparency}}. \bibinfo{pages}{10–19}.
\newblock
\urldef\tempurl%
\url{https://doi.org/10.1145/3287560.3287566}
\showDOI{\tempurl}


\bibitem[\protect\citeauthoryear{Vaccaro, Karahalios, Mulligan, Kluttz, and
  Hirsch}{Vaccaro et~al\mbox{.}}{2019}]%
        {vaccaro2019}
\bibfield{author}{\bibinfo{person}{Kristen Vaccaro}, \bibinfo{person}{Karrie
  Karahalios}, \bibinfo{person}{Deirdre~K. Mulligan}, \bibinfo{person}{Daniel
  Kluttz}, {and} \bibinfo{person}{Tad Hirsch}.}
  \bibinfo{year}{2019}\natexlab{}.
\newblock \showarticletitle{Contestability in Algorithmic Systems}. In
  \bibinfo{booktitle}{\emph{Conference Companion Publication of the 2019
  Conference on Computer Supported Cooperative Work and Social Computing}}.
  \bibinfo{pages}{523--527}.
\newblock
\urldef\tempurl%
\url{https://doi.org/10.1145/3311957.3359435}
\showDOI{\tempurl}


\bibitem[\protect\citeauthoryear{Venkatasubramanian and
  Alfano}{Venkatasubramanian and Alfano}{2020}]%
        {venkatasubramanian2020}
\bibfield{author}{\bibinfo{person}{Suresh Venkatasubramanian} {and}
  \bibinfo{person}{Mark Alfano}.} \bibinfo{year}{2020}\natexlab{}.
\newblock \showarticletitle{The Philosophical Basis of Algorithmic Recourse}.
  In \bibinfo{booktitle}{\emph{Proceedings of the 2020 Conference on Fairness,
  Accountability, and Transparency}}. \bibinfo{pages}{284–293}.
\newblock


\bibitem[\protect\citeauthoryear{Wachter, Mittelstadt, and Floridi}{Wachter
  et~al\mbox{.}}{2017}]%
        {wachter2017}
\bibfield{author}{\bibinfo{person}{Sandra Wachter}, \bibinfo{person}{Brent
  Mittelstadt}, {and} \bibinfo{person}{Luciano Floridi}.}
  \bibinfo{year}{2017}\natexlab{}.
\newblock \showarticletitle{Why a right to explanation of automated
  decision-making does not exist in the General Data Protection Regulation}.
\newblock \bibinfo{journal}{\emph{International Data Privacy Law}}
  \bibinfo{volume}{7}, \bibinfo{number}{2} (\bibinfo{year}{2017}),
  \bibinfo{pages}{76--99}.
\newblock


\bibitem[\protect\citeauthoryear{Wachter, Mittelstadt, and Russell}{Wachter
  et~al\mbox{.}}{2018}]%
        {wachter2018counterfactual}
\bibfield{author}{\bibinfo{person}{Sandra Wachter}, \bibinfo{person}{Brent
  Mittelstadt}, {and} \bibinfo{person}{Chris Russell}.}
  \bibinfo{year}{2018}\natexlab{}.
\newblock \showarticletitle{Counterfactual explanations without opening the
  black box: Automated decisions and the GDPR}.
\newblock \bibinfo{journal}{\emph{Harvard Journal of Law and Technology}}
  \bibinfo{volume}{31} (\bibinfo{year}{2018}).
\newblock


\bibitem[\protect\citeauthoryear{Wang, Yang, Abdul, and Lim}{Wang
  et~al\mbox{.}}{2019}]%
        {wang2019}
\bibfield{author}{\bibinfo{person}{Danding Wang}, \bibinfo{person}{Qian Yang},
  \bibinfo{person}{Ashraf Abdul}, {and} \bibinfo{person}{Brian~Y. Lim}.}
  \bibinfo{year}{2019}\natexlab{}.
\newblock \showarticletitle{Designing theory-driven user-centric explainable
  AI}. In \bibinfo{booktitle}{\emph{Proceedings of the 2019 CHI Conference on
  Human Factors in Computing Systems}}.
\newblock
\urldef\tempurl%
\url{https://doi.org/10.1145/3290605.3300831}
\showDOI{\tempurl}


\bibitem[\protect\citeauthoryear{Warshaw, Taft, and Woodruff}{Warshaw
  et~al\mbox{.}}{2016}]%
        {warshaw2016intuitions}
\bibfield{author}{\bibinfo{person}{Jeffrey Warshaw}, \bibinfo{person}{Nina
  Taft}, {and} \bibinfo{person}{Allison Woodruff}.}
  \bibinfo{year}{2016}\natexlab{}.
\newblock \showarticletitle{Intuitions, Analytics, and Killing Ants: Inference
  Literacy of High School-educated Adults in the US}. In
  \bibinfo{booktitle}{\emph{Proceedings of the Twelfth Symposium on Usable
  Privacy and Security (SOUPS) 2016}}. \bibinfo{pages}{271--285}.
\newblock


\bibitem[\protect\citeauthoryear{West}{West}{2018a}]%
        {west2018brookings}
\bibfield{author}{\bibinfo{person}{Darrell~M. West}.}
  \bibinfo{year}{2018}\natexlab{a}.
\newblock \showarticletitle{Brookings survey finds divided views on artificial
  intelligence for warfare, but support rises if adversaries are developing
  it}.
\newblock \bibinfo{journal}{\emph{Brookings}} (\bibinfo{date}{August}
  \bibinfo{year}{2018}).
\newblock


\bibitem[\protect\citeauthoryear{West}{West}{2018b}]%
        {west2018censored}
\bibfield{author}{\bibinfo{person}{Sarah~Myers West}.}
  \bibinfo{year}{2018}\natexlab{b}.
\newblock \showarticletitle{Censored, suspended, shadowbanned: User
  interpretations of content moderation on social media platforms}.
\newblock \bibinfo{journal}{\emph{New Media \& Society}} \bibinfo{volume}{20},
  \bibinfo{number}{11} (\bibinfo{year}{2018}), \bibinfo{pages}{4366--4383}.
\newblock
\urldef\tempurl%
\url{https://doi.org/10.1177/1461444818773059}
\showDOI{\tempurl}


\bibitem[\protect\citeauthoryear{Woodruff}{Woodruff}{2019}]%
        {woodruff2019}
\bibfield{author}{\bibinfo{person}{Allison Woodruff}.}
  \bibinfo{year}{2019}\natexlab{}.
\newblock \showarticletitle{10 things you should know about algorithmic
  fairness}.
\newblock \bibinfo{journal}{\emph{Interactions}} \bibinfo{volume}{26},
  \bibinfo{number}{4} (\bibinfo{year}{2019}), \bibinfo{pages}{47--51}.
\newblock


\bibitem[\protect\citeauthoryear{Zhang and Dafoe}{Zhang and Dafoe}{2019}]%
        {zhang2019artificial}
\bibfield{author}{\bibinfo{person}{Baobao Zhang} {and} \bibinfo{person}{Allan
  Dafoe}.} \bibinfo{year}{2019}\natexlab{}.
\newblock \showarticletitle{Artificial intelligence: American attitudes and
  trends}.
\newblock \bibinfo{journal}{\emph{Available at SSRN 3312874}}
  (\bibinfo{year}{2019}).
\newblock


\end{thebibliography}
